\documentclass[letterpaper,11pt]{article}
\usepackage{jheppub}
\pdfoutput=1
\usepackage{amsmath,amssymb,amsfonts,bm,amscd,mathtools}
\usepackage{graphicx, caption, subcaption}
\usepackage{multirow}
\usepackage{verbatim}
\usepackage{appendix}
\usepackage{fancybox}
\usepackage{array}
\usepackage{url}
\usepackage{float}

\usepackage{enumitem}

\def\be{\begin{equation}}
\def\ee{\end{equation}}
\def\bea{\begin{eqnarray}}
\def\eea{\end{eqnarray}}

\newcommand{\Dd}{\delta}

\title{Bound on the central charge of CFTs in large dimension}

\author{Abhijit Gadde$^{a}$, Mrunmay Jagadale$^{b}$, Shraiyance Jain$^{a}$ and Trakshu Sharma$^{a}$}
\affiliation{$^{a}$ Department of Theoretical Physics, Tata Institute of Fundamental Research,\\
Mumbai 400005, India.\\
$^{b}$ Walter Burke Institute for Theoretical Physics, California Institute of Technology,\\
 Pasadena, CA 91125, USA.}
\emailAdd{abhijit@theory.tifr.res.in, mjagadal@caltech.edu, shraiyance.jain@tifr.res.in,  trakshu.sharma@tifr.res.in}

\abstract{
  In this paper, we use crossing symmetry and unitarity constraints to put a lower bound on the central charge of conformal field theories in large space-time dimensions $D$. Specifically, we work with the four-point function of identical scalars $\phi$ with scaling dimension $\Delta_{\phi}$, and use a certain class of analytic functionals to show that the OPE coefficient squared $c^2_{\phi \phi T^{\mu \nu}}$ must be exponentially small in $D$. For this to hold, we need to make a mild assumption about the nature of the spectrum below $2\Delta_\phi$. Our argument is robust and can be applied to any OPE coefficient squared $c^2_{\phi \phi O}$ with $\Delta_O< 2\Delta_\phi$. This suggests that conformal field theories in large dimensions (if they exist) must be exponentially close to generalized free field theories.
}

\begin{document}
\maketitle

\section{Introduction}

%\cite{Rattazzi:2008pe,Fitzpatrick:2012yx,Alday:2015ewa,Bissi:2022mrs,Caron-Huot:2020adz,Mazac:2019shk,Caron-Huot:2021enk,Gopakumar:2016wkt,Gopakumar:2018xqi,Mazac:2018ycv,Paulos:2020zxx,Ghosh:2021ruh,Kaviraj:2018tfd,Jafferis:2017zna,El-Showk:2012vjm,Gopakumar:2021dvg,Paulos:2019fkw,El-Showk:2016mxr,Paulos:2021jxx,Dymarsky:2017yzx,Caracciolo:2014cxa,}

The central charge of a conformal field theory is a measure of its degrees of freedom. For a unit-normalized stress tensor $T^{\mu\nu}$, the central charge is inversely proportional to the three-point function coefficient squared $c_{\phi\phi T^{\mu\nu}}^2$ for any operator $\phi$. Hence, it governs the strength of the gravitational coupling in the dual theory in anti-de Sitter space. Therefore, bounding the central charge is important from the point of view of charting not only the space of CFTs but also the landscape of quantum gravity in AdS.

Lower bounds on the central charge of CFTs in two and four dimensions have been computed using numerical bootstrap \cite{Rattazzi:2010gj,Poland:2010wg}.\footnote{See \cite{Simmons-Duffin:2016gjk, Rychkov:2016iqz, Poland:2018epd, Chester:2019wfx} and references therein for an introduction and review of the vast literature on the conformal bootstrap program and \cite{Bissi:2022mrs} and references therein for a focus on analytic methods.}
It is believed that no non-trivial conformal field theory exists in dimensions greater than six\footnote{See \cite{Gadde:2020nwg} for discussion regarding this point.}. Here, a non-trivial CFT means that it is (a) unitary, (b) not free, and (c) contains the stress tensor in its spectrum. One can alternatively formulate this conjecture as:
\begin{itemize}
  \item In $D> 6$, a unitary CFT that is not free must have $c^2_{\phi \phi T_{\mu \nu}}=0$.
\end{itemize}
The advantage of this formulation is that it opens a way of addressing this problem in a quantitative way. In this paper, we will show:
\begin{itemize}
  \item In large $D$, a unitary CFT - with a reasonable condition on the low lying spectrum (this includes not being free) - must have $c^2_{\phi \phi T_{\mu \nu}} < A_\phi e^{-\alpha_\phi D}$. The exponent $\alpha_\phi$ is an ${\cal O}(1)$ number given in equation \eqref{exponent}.
\end{itemize}
The problem of constraining CFTs in large dimensions was considered in \cite{Gadde:2020nwg} by two of the present authors. It was concluded that the unitary CFTs in large dimensions must be exponentially close to generalized free field theories in a certain sense. Happily, our results in this paper concur with the results of \cite{Gadde:2020nwg}. We will comment more on the connection towards the end of section \ref{sectionBoundsonCT}.

The exponential lower bound on the central charge may lead one to naively conclude that there is an exponential hierarchy between the cosmological constant scale and the Planck scale in large dimensions. This is not so. For a $D$-dimensional CFT,
\begin{align}
  M_{\Lambda}^{D-1} G_{N}\sim c^2_{\phi\phi T_{\mu\nu}} = A_\phi e^{-\alpha_\phi D}\qquad \Rightarrow \qquad \frac{M_\Lambda}{M_{P}}\sim e^{-\alpha_\phi}.
\end{align}

We will use the crossing symmetry and unitarity of the four-point function of identical scalars $\phi$ in a large $D$ CFT to put upper bounds on the OPE coefficient $c^2_{\phi \phi T_{\mu \nu}}$. This is accomplished using analytic functional bootstrap \cite{Mazac:2016qev,Qiao:2017lkv, Mazac:2018mdx,Mazac:2018ycv,Mazac:2019shk,Paulos:2019gtx,Kravchuk:2020scc}. In particular, we will use the analytic functionals in \cite{Mazac:2018mdx} that were used to bound certain OPE coefficients in one-dimensional CFTs in the large $\Delta_\phi$ limit. We will review these tools in section \ref{FunctionalBootstrap}. Their application to CFTs in large dimensions is made in section \ref{sectionLD}. The lower bound on the central charge for CFTs in large dimensions is obtained in section \ref{sectionBoundsonCT}. In section \ref{sectionNumericalBounds}, we use the same functionals to obtain approximate numerical bounds for CFTs in large but finite dimensions. The paper contains two appendices that supplement the discussion in the bulk of the paper.

\section{Review of analytic functional bootstrap} \label{FunctionalBootstrap}
Consider the four-point function of identical scalar primary operators $\phi$ of dimension $\Delta_\phi$. This four-point function is fixed by conformal symmetry up to a function of cross-ratios $(u,v)$ as,
\begin{align}
  &\langle \phi(x_{1})\phi(x_{2})\phi(x_{3})\phi(x_{4})\rangle = \frac{1}{|x_{12}x_{34}|^{2\Delta_{\phi}}} G(u,v)\\
  {\rm where}\quad & u\equiv z \bar z =\frac{x_{12}^2 x_{34}^2}{x_{13}^2 x_{24}^2},\, v\equiv (1-z)(1-\bar z) =\frac{x_{14}^2x_{23}^2}{x_{13}^2 x_{24}^2}.\notag
\end{align}
The s-channel OPE expansion i.e. corresponding to $x_{1}\rightarrow x_{2}, x_{3}\rightarrow x_{4} $ is convergent in $z,\bar z\in {\mathbb C}\setminus [1,\infty)$. The t-channel OPE expansion $x_{1}\rightarrow x_{4}, x_{2}\rightarrow x_{3}$ is convergent in $z,\bar z\in {\mathbb C}\setminus (-\infty,0]$. We will consider the correlator in the overlapping region of convergence $(z,\bar z)\in {\cal R}\times {\cal R}$ where ${\cal R}=\mathbb{C}\setminus \{ (-\infty,0] \cup [1,\infty) \}$. The OPE expansions take the form,
\begin{align}
    G(u,v) &=_{s} \sum_{\mathcal{O}\in \phi \times \phi  } c_{\phi \phi \mathcal{O}}^{2} G_{\Delta_{\mathcal{O}},\ell_{\mathcal{O}}}(z,\bar z) \label{sch}  \\
    G(u,v) &=_{t} \frac{(z \bar z)^{\Delta_{\phi}}}{((1-z)(1-\bar z))^{\Delta_{\phi}}} \sum_{\mathcal{O}\in \phi \times \phi  } c_{\phi \phi \mathcal{O}}^{2} G_{\Delta_{\mathcal{O}},\ell_{\mathcal{O}}}(1-z,1-\bar z) \label{tch}
\end{align}
Due to conformal symmetry and permutation symmetry only operators with even spin $\ell$ appear in these expansions. 
The equality of expansions \eqref{sch} and \eqref{tch} is called the crossing equation. It is convenient to express the crossing equation in terms of an elegant sum rule, 
\begin{align}
\sum_{\mathcal{O}\in \phi \times \phi } c_{\phi \phi \mathcal{O}}^{2} F_{\Delta_{\mathcal{O}},\ell_{\mathcal{O}}}^{\Delta_{\phi}}(z,\bar z)&=0 \label{sumrule}\\
{\rm where}\qquad F_{\Delta_{\mathcal{O}},\ell_{\mathcal{O}}}^{\Delta_{\phi}}(z,\bar z) &\equiv \frac{G_{\Delta_{\mathcal{O}},\ell_{\mathcal{O}}}(z,\bar z) }{(z \bar z)^{\Delta_{\phi}}} - \frac{G_{\Delta_{\mathcal{O}},\ell_{\mathcal{O}}}(1-z,1-\bar z) }{((1-z)(1- \bar z))^{\Delta_{\phi}}}.\notag
\end{align}
The functions $F_{\Delta_{\mathcal{O}},\ell_{\mathcal{O}}}^{\Delta_{\phi}}(z,\bar{z})$ are holomorphic in ${\cal R}\times {\cal R}$ and obey 
\begin{align}
  F(z,\bar z) = F(\bar z,z) = - F(1-z,1-\bar z).
\end{align}
Let us call the vector space of such functions ${\cal V}$.
Unitarity implies $c_{\phi \phi \mathcal{O}} \in \mathbb{R} $ and hence $c_{\phi \phi \mathcal{O}}^2 \geq 0$. Therefore, the sum rule sets a positive linear combination of the vectors $F_{\Delta_{\mathcal{O}},\ell_{\mathcal{O}}}^{\Delta_{\phi}}(z,\bar{z})$ to zero. 

Consider a linear functional $\omega$, that is an element of the dual of the vector space. One can act this functional $\omega$ on the sum rule \eqref{sumrule} to get, 
\begin{equation}\label{fsumrule}
\sum_{\mathcal{O}\in \phi \times \phi } c_{\phi \phi \mathcal{O}}^{2} \ \omega \left( F_{\Delta_{\mathcal{O}},\ell_{\mathcal{O}}}^{\Delta_{\phi}}(z,\bar{z}) \right)=0.
\end{equation}
Some simple example of such functionals include evaluation and taking derivatives at a point in $ \mathcal{R}\times \mathcal{R}$. The numerical bootstrap typically uses $\omega$ to be  derivatives at the crossing symmetric point $z = \bar z = \frac{1}{2}$. Note that to get \eqref{fsumrule} from \eqref{sumrule} we have swapped the action of functional $\omega$ with an \emph{infinite} sum over operators appearing in the OPE. Not all functionals satisfy this property. Following \cite{Qiao:2017lkv} we call this property of $\omega$, the swapping condition. Further we want the functionals to be finite on $F_{\Delta,\ell}^{\Delta_{\phi}}(z,\bar{z}) $ with $\Delta, \ell$ satisfying the unitarity bound i.e. $ \Delta \geq \frac{d-2}{2} $ for $\ell=0$ and $ \Delta \geq \ell+d -2 $ for $\ell >0$. We will only consider functionals which satisfy the swapping and finiteness conditions. 

\subsection{Functionals for OPE coefficient maximization}\label{extremal}
\begin{figure}[t]
\begin{center}
\includegraphics[scale=0.6]{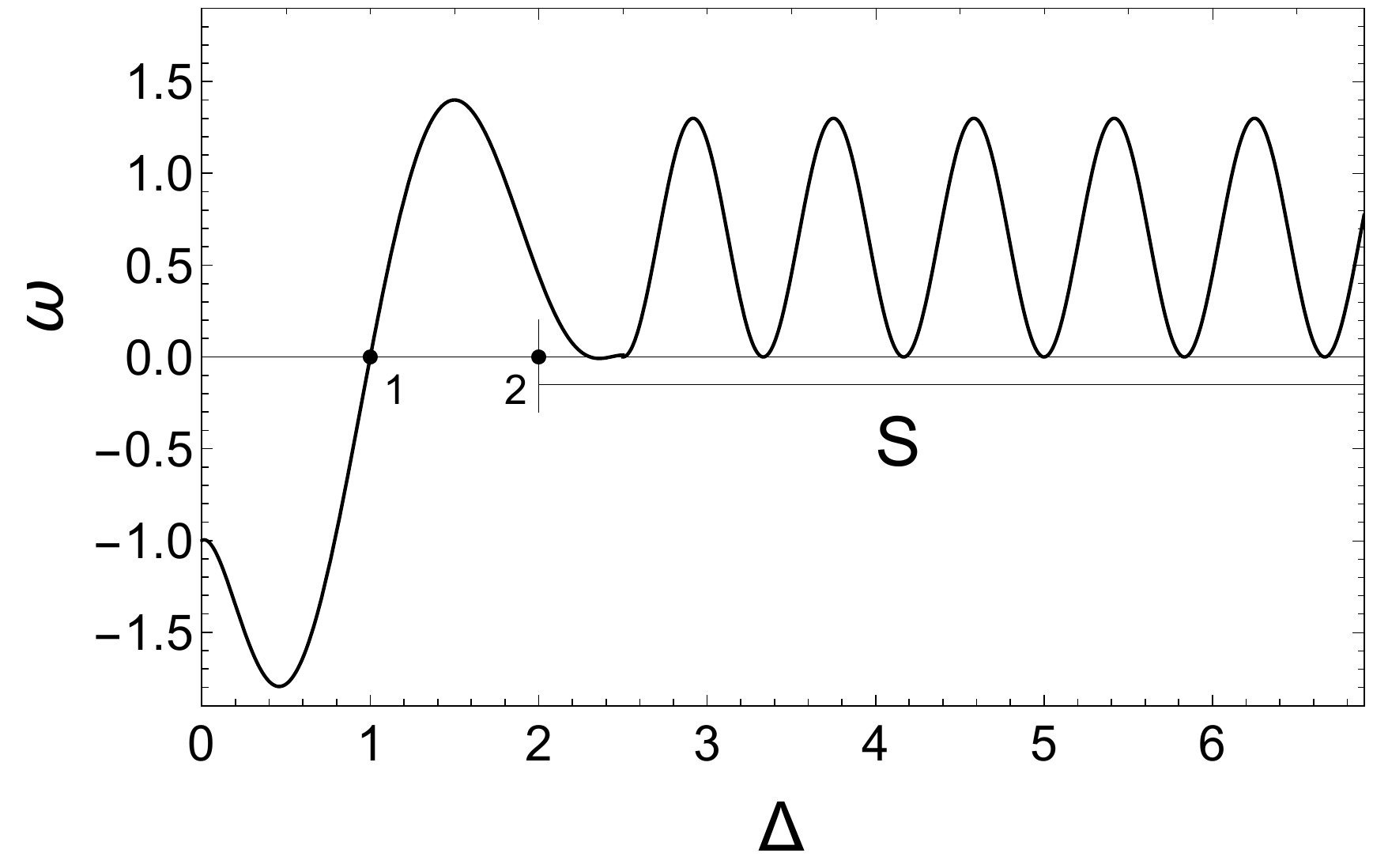}
\caption{Extremal Functional example}
\label{qualextremal}
\end{center}
\end{figure}

In this paper, we will be concerned with obtaining an upper  bound on the OPE coefficient squared $c_{\phi \phi \mathcal{O}_{b}}^{2}$ of a primary operator $\mathcal{O}_{b}\in\phi\times\phi$. Let ${\cal P}$ be the set of all $(\Delta,\ell)$ values where the functional is non-negative and ${\cal S}$ be the set of all CFT operators except for identity ${\bf 1}$ and ${\cal O}_b$.
The sum rule constraining CFT data is, 
\begin{equation}
F_{\bf 1}^{\Delta_{\phi}} (z,\bar z) + c_{\phi \phi \mathcal{O}_{b}}^{2} F_{\Delta_{b},\ell_{b}}^{\Delta_{\phi}}  (z,\bar z)+ \sum_{(\Delta,\ell)  \in {\cal S}} c_{\phi \phi \mathcal{O}}^{2} F_{\Delta,\ell}^{\Delta_{\phi}} (z,\bar z) =0 
\end{equation}
The action of a functional $\omega$ on the sum rule is, 
\begin{equation}
\omega({\bf 1}) + c_{\phi \phi \mathcal{O}_{b}}^{2} \omega(\Delta_{b},\ell_{b}) + \sum_{{\cal S}} c_{\phi \phi \mathcal{O}}^{2} \omega(\Delta,\ell)=0. 
\end{equation}
Therefore, the OPE coefficient $c_{\phi \phi \mathcal{O}_{b}}^{2}$ can be expressed as 
\begin{equation} 
c_{\phi \phi \mathcal{O}_{b}}^{2} = -\frac{\omega({\bf 1})}{ \omega(\Delta_{b},\ell_{b})} - \frac{\sum_{\cal S} c_{\phi \phi \mathcal{O}}^{2} \omega(\Delta,\ell)}{\omega(\Delta_{b},\ell_{b})}
\end{equation}
At this point, it is easy to see that we can obtain an upper bound on $c_{\phi \phi \mathcal{O}_{b}}^{2}$ by constructing a functional that satisfies,
\begin{align}\label{extcondw}
\omega({\bf 1}) &<0, &  \omega(\Delta_{b},\ell_{b}) &>0,  & {\cal S} \subset {\cal P}.
\end{align}
Existence of such functional would give us the bound,
\begin{equation}
c_{\phi \phi \mathcal{O}_{b}}^{2} \leq \frac{-\omega({\bf 1})}{ \omega(\Delta_{b},\ell_{b})}.
\end{equation} 
Moreover this inequality is saturated when  $ \omega(\Delta,\ell) = 0,  \,\, \forall (\Delta,\ell) \in {\cal S}$. Such functionals are called extremal functionals. A cartoon of an extremal functional for one dimensional CFT is given in figure \ref{qualextremal}, where  $\Delta_{b}$ is taken to lie between $1$ and $2$ and $S$ is $\{ \Delta : \Delta > 2 \}$. The set ${\cal S}$ is precisely the set of double zeros of the functional.

\subsection{Analytic functionals for $z=\bar z$}
In this section we will consider analytical functionals that act on the functions restricted to the locus $z=\bar z$.  These functionals were first constructed in \cite{Mazac:2018mdx} for one dimensional CFTs. It is straightforward to repurpose them as functionals for CFTs in general dimensions but acting only on the specialization  $z=\bar z$. Let $\tilde {\cal V}$ be the vector space of functions $F(z)$ that is holomorphic in ${\cal R}$ and obey $F(z)=-F(1-z)$. After specializing a function in ${\cal V}$ to $z=\bar z$, we precisely get a function in $\tilde {\cal V}$.

The authors of \cite{Mazac:2018mdx} consider a class of  functionals acting on $\tilde {\cal V}$ given by the integral of the discontinuity along the branch cut $[1,\infty)$ weighted by a kernel $h(z)$.
\be \label{1dfunc}
\omega(F) = \frac{1}{2\pi i} \int_{1}^{\infty} d z \hspace{0.25cm} h(z)  \textrm{Disc}[F(z)] =\frac{1}{2\pi i} \int_{-\infty}^{0} d z\hspace{0.25cm} h(1-z) \textrm{Disc}[F(z)]  .
\ee 
Here $\textrm{Disc}[F(z)] = \lim_{\epsilon \rightarrow 0^{+}} F(z + i \epsilon ) - F(z - i \epsilon ) $ is the discontinuity along the branch cut. In the second equality we have done the change of variables from $z \rightarrow 1-z$ and used $F(1-z) =-F(z)$. The kernel $h(z)$ is analytic on $\mathbb{C}\setminus{(-\infty,1)}$ with possible branch cuts at $(-\infty,0]$ and $[0,1]$. Without loss of generality, we can assume $h(z) \in \mathbb{R}$ for $z \in (1,\infty)$. This implies $h^*(z)=h(z^*)$. The kernel $h(z)$ satisfies the properties,
\begin{enumerate}
\item $h(z)$ is analytic away from possible poles or branch points at $z = 0, \ 1 $ and $\infty$. 
\item $h(z)$ is bounded by $ A_{1} |z|^{-1 - \epsilon_{1}}$ for some $A_1, \epsilon_1 > 0$ as $z \rightarrow \infty$. 
\item The discontinuity of $h(z)$ around $z = 1$ is bounded by $ A_{2} |z-1|^{2 \Delta_{\phi}-1 + \epsilon_{2}}$
for some $A_2, \epsilon_2 > 0$ as $z \rightarrow 1$.
\end{enumerate}
The second and the third properties follow from finiteness and swapping conditions respectively. See \cite{Mazac:2018mdx, Qiao:2017lkv} for details regarding this point. The action of the functional in \eqref{1dfunc} on the function $F_{\Delta,\ell}^{\Delta_{\phi}}(z,z)$ is given by 
\be \label{1dfuncG}
\omega(\Delta,\ell) = \frac{1}{2\pi i} \int_{-\infty}^{0} d z \hspace{0.25cm} h(1-z) \textrm{Disc}\left[ \frac{G_{\Delta,\ell}(z)}{z^{2\Delta_{\phi}}} - \frac{G_{\Delta,\ell}(1-z)}{(1-z)^{2\Delta_{\phi}}} \right]
\ee 
Here, we use $G_{\Delta,\ell}(z)=G_{\Delta,\ell}(z,z)$ for the conformal block specialized to $z=\bar z$. After certain contour manipulations detailed in appendix \ref{functional-simple}, the functional reduces to,
\begin{equation}\label{simplifiedw}
 \omega(\Delta,\ell) = \mathfrak{g}(\Delta,\ell) -  \mathrm{Im}\left[ e^{i \pi (\Delta-2\Delta_{\phi})} \mathfrak{f}(\Delta,\ell) \right],
\end{equation} 
where, $\mathfrak{f}$ and $\mathfrak{g}$ are given by,
\begin{align}
\mathfrak{f}(\Delta,\ell) &=  \int_{-\infty}^{0}d z \hspace{0.25cm} f(z) \frac{\hat{G}_{\Delta,\ell}(z)}{|z|^{2\Delta_{\phi}}} \label{funcf} \\
\mathfrak{g}(\Delta,\ell) &=  \int_{-\infty}^{0}d z \hspace{0.25cm} (1-z)^{2\Delta_{\phi}-2} g\left(\frac{z}{z-1}\right) \frac{\hat{G}_{\Delta,\ell}(z)}{|z|^{2\Delta_{\phi}}} =   \int_{0}^{1} d z \hspace{0.25cm} g(z) \frac{G_{\Delta,\ell}(z)}{z^{2\Delta_{\phi}}} \ . \label{funcg}
\end{align}
Here we have introduced\footnote{Our definition of $f(z)$ differs from the one used in \cite{Mazac:2018mdx} by a factor of $i$. That is the reason ${\rm Im}[f(z)]$ appears in our gluing condition \eqref{gluingcond} rather than ${\rm Re}[f(z)]$.}
\begin{align}
  \hat{G}_{\Delta,\ell}(z) &= |G_{\Delta,\ell}(z+ i \epsilon)|    & \text{ for } & z \in (-\infty,0) \\
  f(z) &=   \frac{ h(z)- h(1-z)}{\pi } & \text{ for } & \mathrm{Im}(z) > 0, \label{1df} \\
  g(z) &= -   \frac{  \textrm{Disc}\left[h(z)  \right]}{2 \pi i} & \text{ for } & z \in(0,1)\label{1dg}
\end{align}
The function $g$ is analytically continued in the $z$ variable and the function $f$ is continued to the region with $ \mathrm{Im}(z) < 0$ via $f(z) \equiv - f(1-z)$. With this analytic continuation, we have $f^{*}(z)= f(z^*)$. The definitions of the kernels $f$ and $g$ in terms of the kernel $h$ imply that they obey the following relation called the gluing condition,
\begin{equation}\label{gluingcond}
  \mathrm{Im}[f(z)] + g(z)+ g(1-z) = 0 \hspace{1cm} \text{ for } z \in(0,1).
\end{equation}

This manipulation is valid for $\Delta > \Delta_{c}$, where $\Delta_{c}$ is some positive scaling dimension. This is because although the functional $\omega(\Delta)$ is always finite by definition, the individual integrals in equations \eqref{funcf} and \eqref{funcg} can diverge near $z=0$. Since the conformal block goes like $|z|^{\Delta}$ as $z \rightarrow 0$, the integrals \eqref{funcf} and \eqref{funcg} are convergent for $\Delta$ greater than certain $\Delta_c$. The precise value of $\Delta_c$ depends on the behavior of the kernel $h(z)$ near $z=0$.
Outside this range i.e. for $\Delta<\Delta_c$, we need to resort to the manifestly finite expression of the functional \eqref{1dfuncG} for evaluation.

Since $\textrm{Disc}[h(z)]$ is purely imaginary and $G_{\Delta,\ell}(z)$ is real for $z \in (0,1)$, this means $\mathfrak{g}(\Delta,\ell)$ is real. However $h(z+ i \epsilon)$ generically has both real and imaginary parts. So $\mathfrak{f}(\Delta,\ell)$ can be written as  $\mathfrak{f}(\Delta,\ell) = i \mathfrak{r}(\Delta,\ell)e^{i \pi \gamma(\Delta,\ell)}$ with $\mathfrak{r}(\Delta,\ell)\in \mathbb{R}_{+}$ and $\gamma(\Delta,\ell)\in \mathbb{R}$. With this, the functional \eqref{simplifiedw} becomes,
\begin{equation}\label{simplifiedfunc}
\omega(\Delta,\ell) = \mathfrak{g}(\Delta,\ell) - \mathfrak{r}(\Delta,\ell) \cos(\pi(\Delta-2\Delta_{\phi} + \gamma(\Delta,\ell))). 
\end{equation} 
If the choice of $h$ is such that $\mathfrak{g}$, $\mathfrak{r}$, $\mathfrak{\gamma}$ are slowly varying compared to the oscillations of the cosine in above equation and further if $ \mathfrak{r}(\Delta,\ell) = \mathfrak{g}(\Delta,\ell)$, %$\left(\text{ or } |f(z)| \approx (1-z)^{2\Delta_{\phi}-2} \vert g\left(\frac{z}{z-1}\right)\vert \right)$ for large enough $\Delta$, 
then one gets an extremal functional which has double zeros at  
\begin{equation}\label{double-zero}
\Delta_{n} = 2 \Delta_{\phi} +  2n - \gamma(\Delta_{n},\ell) \hspace{1.5cm} n \in \mathbb{Z}_{\geq0} .
\end{equation} 

The functional, in particular the integrals $\mathfrak{f}(\Delta,\ell)$ and $\mathfrak{g}(\Delta,\ell)$ are difficult to compute analytically in general. If the conformal dimensions $\Delta_\phi$ and $\Delta$ both are taken to be large then these integrals can be performed via saddle point approximation. So at this stage, we will take the limit of large $D$. This will also set all the conformal dimensions to be large, thanks to the unitarity bound. 

If we want the saddle points of $\mathfrak{f}(\Delta,\ell)$ and $\mathfrak{g}(\Delta,\ell)$ to be universal and not depend on the kernel $f$ and $g$, and if we further want $\mathfrak{f}(\Delta,\ell)$ and $\mathfrak{g}(\Delta,\ell)$ to be of the same order (which is necessary to get the double zeroes \eqref{double-zero}) then we need to take 
\begin{align}\label{fgscaling}
  &f(z) \sim {\cal O}(1) \qquad \qquad \qquad \qquad \qquad \qquad \qquad \quad\,\,\textrm{for Im}(z)>0, \notag \\
  &g(z) = (1-z)^{2 \Delta_{\phi}} \tilde{g}(z),\,\quad {\rm with}\,\,\, {\tilde g}(z)\sim{\cal O}(1)  \qquad \textrm{for } z \in (0,1). 
\end{align}
In the limit of large $\Delta, \Delta_{\phi}$ and with the scaling of kernels $f(z)$ and $g(z)$ given above, it is easy to see that the integrals for ${\mathfrak f}$ and ${\mathfrak g}$ are convergent for $\Delta$ below some $\Delta_c$. If $f(z)={\cal O}(z^{-1+\epsilon})$ near $z=0$ then $\Delta_c=2\Delta_\phi$. We will assume that $f(z)$ obeys this property.
Therefore, the computation of the functional $\omega(\Delta)$ can be divided into three regions, depending on the method of computation, namely (I) $\omega(\Delta > 2 \Delta_{\phi})$, (II) $\omega(0< \Delta < 2 \Delta_{\phi})$ and (III) $\omega({\bf 1})$.

\subsection{Computing the functional in large $D$}\label{sectionLD}
In this section we will compute the functional $\omega(\Delta,\ell)$ in the large $D$ limit. For this we will need the conformal block in the large $D$ limit.  These blocks were first computed in \cite{Fitzpatrick:2013sya}. They gave an explicit expression in terms of the hypergeometric function,
\begin{equation} 
 {G}_{\Delta,\ell} (y_{+},y_{-}) = \frac{2^\Delta}{\sqrt{y_{-}-y_{+}}} A_{\Delta}(y_{+})A_{1-\ell}(y_{-})     
\end{equation}
with, 
\begin{align}\label{Afunction}
A_{\beta}(x) &= x^{\frac{\beta}{2}} \prescript{}{2}{F}_{1}\left(  \frac{\beta}{2}, \frac{\beta-1}{2}, \beta-\frac{D}{2}+1, x   \right)  \ , \qquad y_{\pm}  = \frac{z \bar z}{(1 \pm \sqrt{(1-z)(1-\bar z)})^{2}}. 
\end{align}
One immediate observation is that on the $z=\bar z$ slice,  $y_-=1$. Hence, the spin dependance of the block trivializes as we get $A_{1-\ell}(1)=1$. We will drop the spin label from now on as the functional $\omega(\Delta,\ell)$ does not depend on $\ell$ in the large $D$ limit.
We are interested in scaling all $\Delta=\delta D$. In this limit the large $D$ block simplifies even further on $z=\bar z$ locus. We get
\begin{equation}\label{LDblockSP}
  G_{\delta D}(z) =   \Big(1-\left( \frac{z}{2-z}\right)^{2}\Big)^{-\frac12} \ \mathsf{v}_{\delta}(z) \ e^{D \hspace{.02cm} \mathsf{q}_{\delta}(z)}.
  \end{equation}
The functions $\mathsf{v}_\delta(z)$ and $\mathsf{q}_\delta(z)$ are given in equation \eqref{qv}. 
The saddle point in integrals \eqref{funcf} and  \eqref{funcg} come from extremizing $G_{\delta D}(z)/z^{2\delta_\phi D}$ with respect to $z$ in the large $D$ limit. This is computed in appendix \ref{blocksaddlepoint}. We get,
\begin{equation}\label{saddlez}
  z_*(\delta) =    \frac{(2 \delta_{\phi} - \delta )( 2 \delta_{\phi} +\delta -1 )}{\delta_{\phi}(4 \delta_{\phi}-1)}.
\end{equation}
No we will evaluate the functional in the three regions mentioned earlier.\\

\noindent {\bf(I)} $\delta>2\delta_\phi$: 
For this range of $\delta$ the saddle point $z_*\in (-\infty,0)$. The functional is 
\begin{equation}\label{dg2dp}
\omega(\Delta > 2 \Delta_{\phi}) = \mu(\delta_{\phi},z_*) \Big [ (1-z_*)^{-2} \tilde{g}\left( \frac{z_*}{z_*-1} \right) - \cos(\pi (\Delta-2\Delta_{\phi}+ \gamma(z_*))) \ |f(z_*)| \Big]
\end{equation}
where, $\mu(\delta_{\phi},z_*)$ is a positive pre-factor given below. It is independent of $f (z)$ and  $\tilde{g}(z)$.
\begin{equation}
\mu(\delta_{\phi},z_*)=\frac{\hat{G}_{\delta D}(z_{*})}{|z_{*}|^{2\delta_{\phi}D}} \sqrt{\frac{2 \pi }{ - \delta_{\phi}D {\mathsf q}_\delta''( z_{*} )}} .
\end{equation}
The phase $\gamma(z)$ is defined by $f(z+ i 0^{+}) = i |f(z)| e^{i \pi \gamma(z)}$ and is of $\mathcal{O}(1)$. It is clear from the expression that  the functional $\omega(\Delta)$ for $\Delta \geq 2 \Delta_{\phi} $ is oscillating because of the cosine factor. Since we want positive functional for $\Delta> 2 \Delta_{\phi}$, this means the first term in the square brackets should be greater  than or equal to the second, 
\begin{equation}\label{extremalcond}
  (1-z)^{-2} \tilde{g}\left( \frac{z}{z-1} \right) \geq |f(z)| \qquad \textrm{for } z \in (-\infty,0).
\end{equation}
To get an extremal functional we will require this inequality be saturated. Extremality along with the gluing condition \eqref{gluingcond} yields the constraint on the kernel $f(z)$,
\begin{align}\label{realf}
  {\rm Im}(f(z))=z^{2\Delta_\phi -2}|f(1/z)|+(1-z)^{2\Delta_\phi -2}|f(1/(1-z))|=0\qquad \qquad z\in (0,1).
\end{align}
In the second equality we use the fact that $\Delta_\phi$ is large, $z\in (0,1)$ and $f(z)\sim {\cal O}(1)$.

The functional takes the form
\begin{align}\label{doublez}
  \omega(\Delta>2\Delta_\phi)=2\mu(\delta_\phi,z_*)|f(z_*)|\sin^2\Big(\frac{\pi}{2} (\Delta-2\Delta_{\phi}+ \gamma(z_*))\Big).
\end{align}
It has double zeroes at $\Delta=2\Delta_{\phi}+2n-\gamma(z_*)$.\\

\noindent {\bf (II)} $0<\delta<2\delta_\phi$:
For this range of $\delta$ the integrals \eqref{funcf} and \eqref{funcg} are divergent so we have to use the original definition of the functional \eqref{1dfuncG} to evaluate it. This is done in appendix \ref{blocksaddlepoint}. The result is,
\begin{equation}\label{dl2dp}
\omega(0< \Delta < 2 \Delta_{\phi}) = \mathrm{Re} \left(f(z_{*}) \right) \frac{G_{\delta D}(z_{*})}{z_{*}^{2\delta_{\phi}D}} \sqrt{\frac{2 \pi }{ \delta_{\phi} D {\mathsf q}_\delta''( z_{*})}}
\end{equation}
Here $z_*$ is the one defined in \eqref{saddlez}. Notice that the value of the functional scale exponentially with $D$.\\

\noindent {\bf (III)} Identity operator ${\bf 1}$:
This functional is also evaluated in appendix \ref{blocksaddlepoint}. The result is,
\begin{equation}\label{identity}
\omega({\bf 1}) = - \int_{-\infty}^{0} dz |f(z)|.
\end{equation}

Recall the conditions \eqref{extcondw} on $\omega$ to get a bound on the OPE coefficient. 
It is clear from \eqref{identity} that $\omega({\bf 1})<0$ is already obeyed.  To impose the condition $\omega(\Delta_b,\ell)>0$, we need to impose ${\rm Re}(f(z))>0$ for $z\in (0,1)$. As the functional is non-negative for $\delta>2\delta_\phi$, the set of operators with quantum numbers $\{(\Delta,\ell): \Delta >2\Delta_\phi\}$ is definitely contained in ${\cal P}$. If the rest of the CFT operators also lie in the set ${\cal P}$  then the bound on the OPE coefficient squared for $\Delta_b\leq 2\Delta_\phi$ is
\begin{align}\label{1Dbounds}
  c^2_{\phi \phi O_b} &\leq - \frac{\omega({\bf 1})}{\omega(\Delta_b)} = \frac{\int_{-\infty}^{0} dz |f(z)|}{\mathrm{Re}(f(z_{b}))}\Big(
  \sqrt{\frac{\delta_\phi D}{2\pi}}
  \frac{z_{b}^{2\delta_{\phi} D} \sqrt{  {\mathsf q}_\delta''( z_{b} )}}{ G_{\delta_b D}(z_{b}) },\Big)\qquad z_b=z_*(\delta_b).
  \end{align}
The expression inside the big parenthesis is independent of $f(z)$. It is evaluated explicitly in appendix \ref{blocksaddlepoint}. Let us collect all the properties of $f(z)$. 
\begin{enumerate}
\item It is analytic in ${\mathbb C}\setminus (-\infty,\infty)$ with possible singularities only at $0,1$ and $\infty$.
\item $f(z)=-f(1-z)$. This follows from the definition \eqref{1df} of $f(z)$.
\item Im$\left[f(z)\right]=0$ for $z \in (0,1)$. This is derived in  \eqref{realf}.
\item $f(z)=\mathcal{O} \left( z^{-1+\epsilon} \right)$ near $z=0$. This is needed for $\Delta_c=2\Delta_\phi$
\item $f(z)=\mathcal{O} \left( z^{-1-\epsilon} \right)$ near  $z  = \infty$. This follows from decay property of $h(z)$.
\end{enumerate}
If we show the existence of $f(z)$ satisfying above properties then the bound \eqref{1Dbounds} holds. As shown in appendix \ref{blocksaddlepoint},  the bound takes the form $c^2_{\phi \phi O_b}< A(\delta_b,\delta_\phi) e^{-\alpha(\delta_b,\delta_\phi) D}$ with $\alpha(\delta_b,\delta_\phi)>0$ for $\delta_b<2\delta_\phi$. The explicit expressions for $A(\delta_b,\delta_\phi)$ and $\alpha(\delta_b,\delta_\phi)$ are cumbersome and are give in appendix \ref{blocksaddlepoint}.
Note that exponent $\alpha$ is robust and  does not depend on $f(z)$. However, the set ${\cal P}$ of the quantum numbers $(\Delta,\ell)$ where the functional $\omega$ is non-negative depends crucially on $f(z)$. 

\section{Bounds on Central Charge in Large D CFTs}\label{sectionBoundsonCT}
In this section, we will apply the technology of section \ref{FunctionalBootstrap} to obtain lower bound on the central charge $C_T$ of a CFT in large $D$.

When the stress tensor of the CFT is canonically normalized i.e. when it obeys the Ward identity
\begin{align}
  \partial_\mu \langle T^{\mu \nu}\phi(x_1)\ldots \phi(x_n)\rangle = -\sum_i \delta(x-x_i)\langle\phi(x_1)\ldots\partial^\mu \phi(x_i)\ldots \phi(x_n)\rangle,
\end{align}
the central charge governs its two point function.
\begin{align}
  \langle T^{\mu\nu}(x)T^{\lambda\sigma}(0)\rangle=\frac{C_T}{S_d^2}\frac{1}{x^{2d}}\Big(\frac12 (I^{\mu \lambda}I^{\nu\sigma}+I^{\mu \sigma}I^{\nu \lambda})-\frac1D \delta^{\mu\nu}\delta^{\lambda\sigma}\Big).
\end{align}
Here $I^{\mu\nu}(x)\equiv \delta^{\mu\nu}-2x^\mu x^\nu/x^2$ and $S_d=2\pi^{D/2}/\Gamma(D/2)$ is the volume of the $D-1$ dimensional sphere. With this definition of the central charge, its value for a single free degree of freedom is
\begin{itemize}
  \item $C_T^{\rm scalar}=D/(D-1)$
  \item $C_T^{\rm fermion}=D/2$
  \item $C_T^{\rm (D-2)/2 \,form}=D^2/2$.
\end{itemize}
There is $1$ degree of freedom for a free scalar field and this value for a free fermion field and a free $(D-2)/2$-form are $2^{D/2}$ and $\Gamma(D-2)/\Gamma^2(D/2)$ respectively.

\begin{figure}[t]
  \begin{center}
    \includegraphics[scale=0.5]{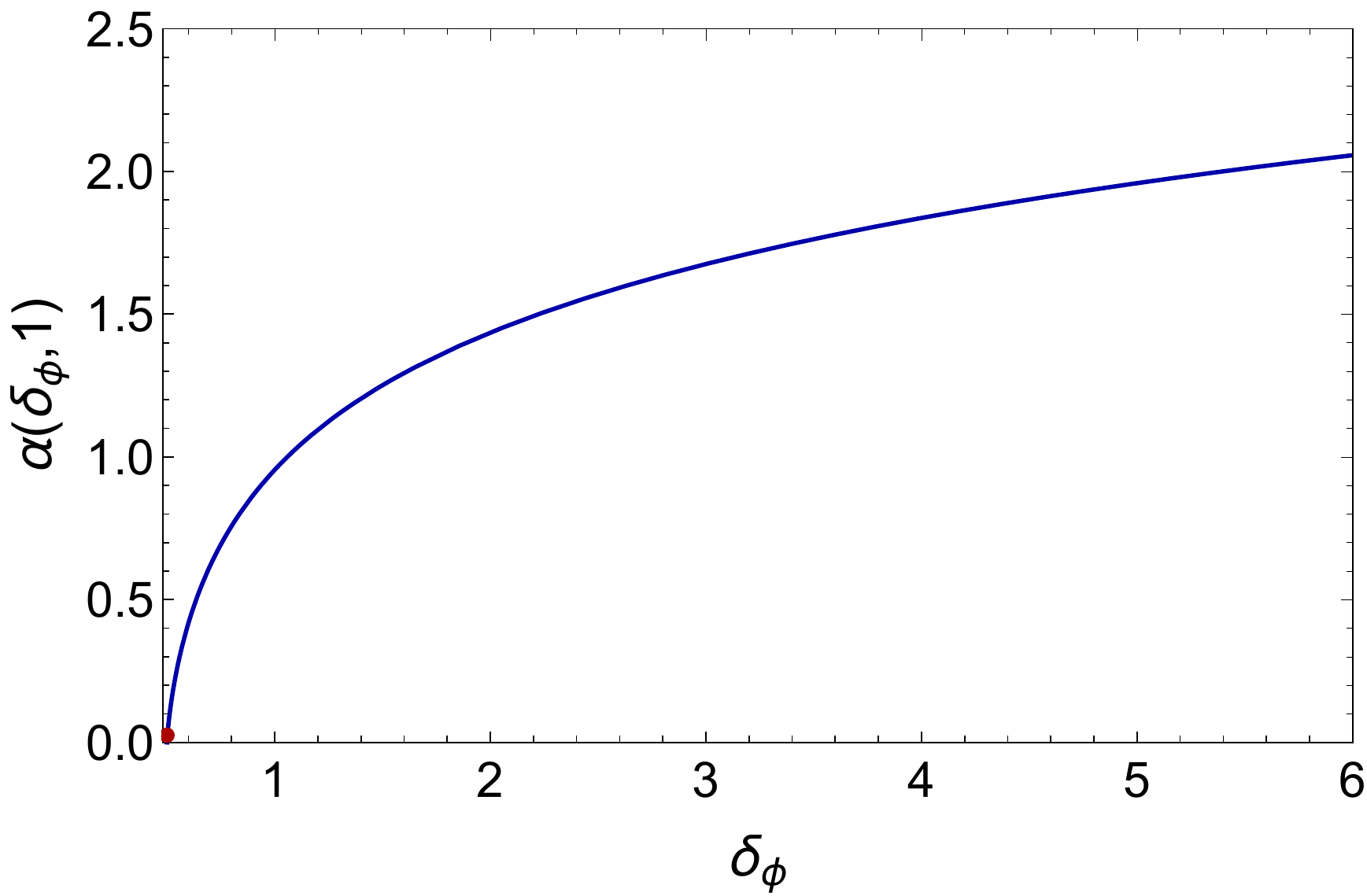}
  \end{center}
    \caption{The function $\alpha(\delta_\phi,1)$ is plotted against $\delta_{\phi}$. It is always greater than zero for $\delta_{\phi}>1/2$.}\label{chifigure} 
  \end{figure}

In order to apply the analytic bounds for the stress tensor we must first normalize its two point function to unity. With the new normalization, $C_T$ appears in the OPE coefficient squared $c^2_{\phi\phi T^{\mu\nu}}$ as 
\begin{align}\label{ct}
  c^2_{\phi\phi T^{\mu\nu}}=\frac{1}{C_T}\Big(\frac{\Delta_\phi D}{(D-1)}\Big)^2.
\end{align}
If we now apply the bound computed in section \ref{FunctionalBootstrap}, we get
\begin{align}\label{Ctbound}
  C_T> \Big(\frac{\Delta_\phi D}{(D-1)}\Big)^2\frac{1}{A(\delta_\phi,1)} e^{\alpha(\delta_\phi,1) D}.
\end{align}
Here we have set $\delta_b=1$ because the role of $O_b$ is played by $T^{\mu\nu}$ and because its conformal dimension is $D$, $\delta_b=1$.  For $\delta_b=1$, the expression for $\alpha$ simplifies.
\begin{align}\label{exponent}
  \alpha(\delta_\phi,1)=\log\left(\frac{2(2\delta_{\phi}-1)}{\sqrt{4 \delta_{\phi}-1} }\left( \frac{4 \delta_{\phi}-1}{2(2\delta_{\phi}-1)} \right)^{2 \delta_{\phi}}\right).
\end{align}
It is easy to see that $\alpha>0$ for all values of $\delta_\phi$ in the unitary range i.e. for $\delta_\phi>1/2$. 
The plot of $\alpha(\delta_\phi,1)$ against $\delta_\phi$ is given in figure \ref{chifigure}.
The function $A(\delta_\phi,1)$ depends on the details of the function $f(z)$. Recall that this bound applies if the CFT operators ${\cal S}\subset {\cal P}$. The set ${\cal P}$ depends on $f(z)$.

By construction the functional $\omega(\Delta,\ell)$ is non-negative for $\Delta \geq 2 \Delta_{\phi}$ (for all $\ell$). The sign of the functional $\omega(\Delta,\ell)$ for $\Delta<2\Delta_{\phi}$ depends only on the sign of $\mathrm{Re} \left(f \left(z_{b} \right)\right) $ and hence can be positive or negative depending on the choice of kernel $f(z)$. Ideally one would want $\mathrm{Re} \left(f \left(z_{b} \right)\right) \geq 0$ for all operators (except identity) satisfying unitarity bounds with $\Delta<2 \Delta_{\phi}$.
This would show that  the central charge is exponentially large in $D$ in \emph{any} large dimensional conformal field theory except for the free theory. We will now show that this condition,
\begin{equation}\label{positivecond}
\textrm{Re} \left(f \left(z_{b} \right)\right) \geq 0 \qquad \qquad \frac{1}{2} \leq \delta < 2 \delta_{\phi},
\end{equation}
is impossible to achieve with the kind of functionals we are using i.e. \eqref{1dfuncG}. To see this, notice that the kernel $f(z)$ is necessarily anti-symmetric around $z=1/2$ (see equation \eqref{1df}). This means $f(1/2)=0$, and the Re$\left(f(1/2 \pm a) \right)$ are of opposite sign for any $a$. The functional $\omega(\Delta,\ell)$ for $\Delta<2 \Delta_{\phi}$ is proportional to $\textrm{Re} \left(f \left(z_{b} \right)\right)$ (times a positive factor), and of opposite sign for $z_{b}=1/2 \pm a$. The functional vanishes for $z_b=1/2$. Solving the condition $z_*(\delta_b)=1/2$, we get a critical value of $\delta_b$ where the functional must vanish,\footnote{The other solution $\frac{1}{2} \left(1-\sqrt{2 \delta \phi -1} \sqrt{4 \delta \phi -1}\right)$ lies outside the unitary region so we ignore it.}
\begin{equation}\label{zeropoint}
\delta_{\rm crit} = \frac{1}{2} \left(1+\sqrt{2 \delta \phi -1} \sqrt{4 \delta \phi -1}\right).
\end{equation}
This means that the functional can never be of the same signature in the whole range $1/2 \leq \delta<2\delta_{\phi}$. Its sign has to change at $\delta_{\rm crit}$. This means we can not satisfy \eqref{positivecond}. 
With this analysis we also conclude that we do not get any bound on the central charge for $\delta_\phi=3/4$. This is because for this value of $\delta_\phi$, $\delta_{\rm crit}=1$. As the functional vanishes for $\delta=1$, the central charge is unbounded from below. For other values of $\delta_\phi$ we can certainly obtain exponentially large lower bound on the central charge given the set of CFT operators ${\cal S}\subset {\cal P}$. We will characterize the set ${\cal P}$ below. As discussed above, the operators with $\delta\geq 2\delta_\phi$ belong to ${\cal P}$ so we will only be concerned with characterizing ${\cal P}$ of  operators with $\delta<2\delta_\phi$.
\begin{figure}[t]
  \begin{subfigure}{.5\textwidth}
    \centering
    \includegraphics[scale=0.4]{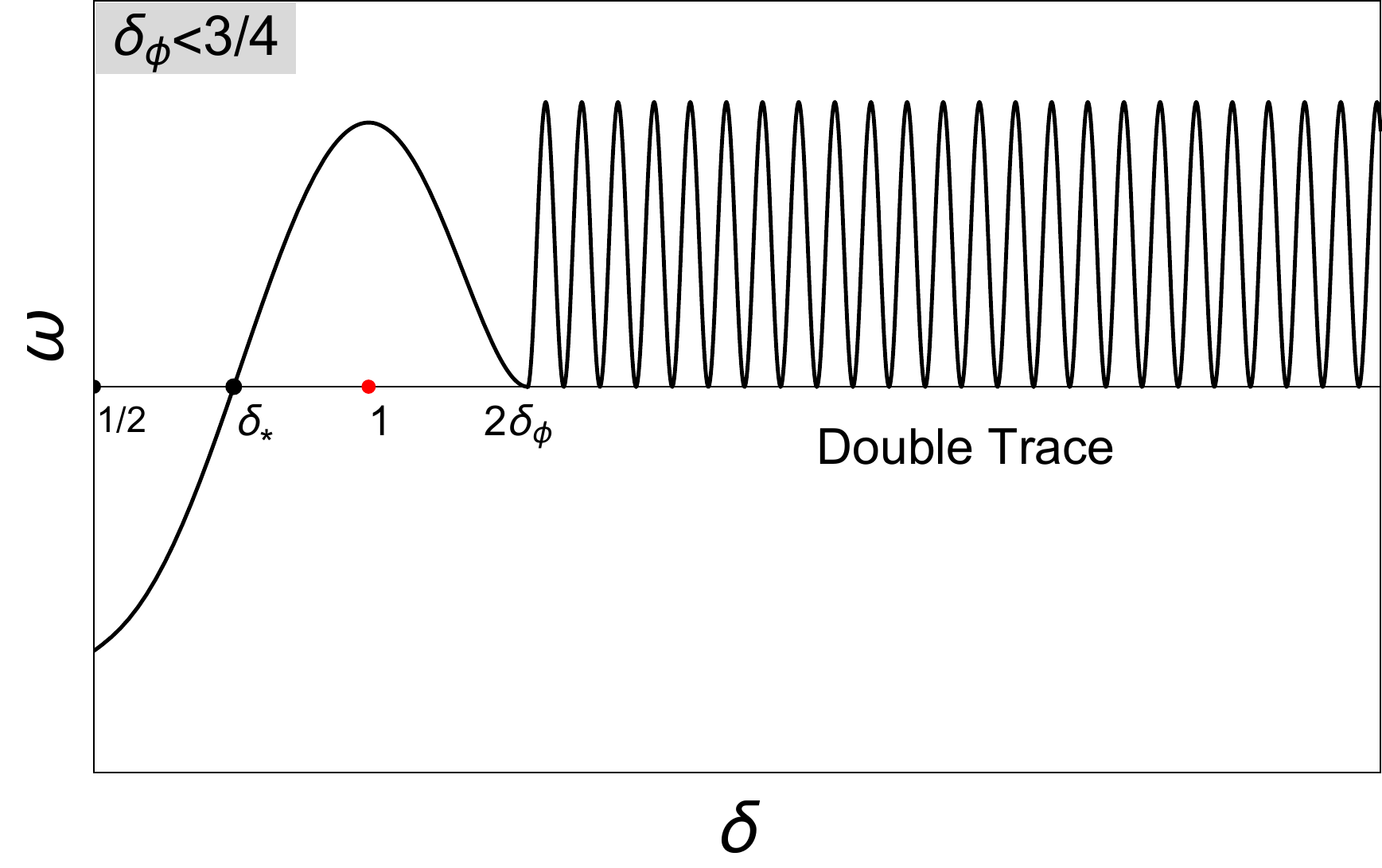}
    \label{lessthanthreequarters}
  \end{subfigure}%
  \begin{subfigure}{.5\textwidth}
    \centering
  \includegraphics[scale=0.4]{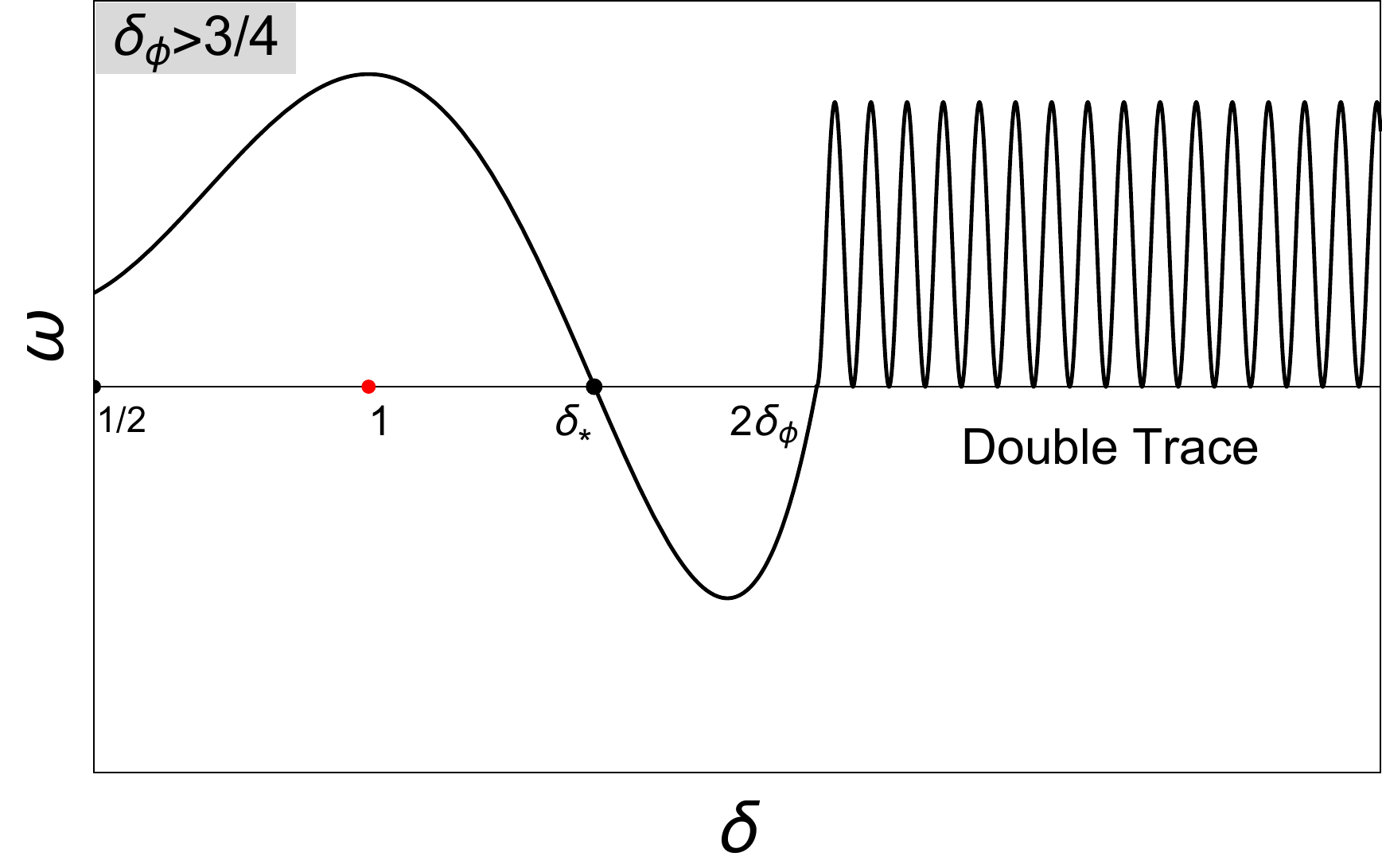}
    \label{greaterthanthreequarters}
  \end{subfigure}
  \caption{Schematic plots of functional $\omega(\delta,\ell)$  with kernel $f(z)=f_{\rm opt}(z)$, for $\delta_{\phi}<3/4$ (left) and $\delta_{\phi}>3/4$ (right).  Functional at stress tensor i.e. with $\delta=1$, is positive and functional at identity (not shown) is negative, in both cases. The $\ell$ dependence is not shown because the signature of  $\omega(\delta,\ell)$ is independent of $\ell$.} 
  \label{lessthangreaterthanthreequarters}
  \end{figure}

\subsection*{Case I: $\delta_\phi<3/4$}
In this case, the $\omega(\Delta)$ is positive for $\Delta> \delta_{\rm crit}D$. The schematic plot of the functional is shown in figure \ref{lessthangreaterthanthreequarters} on the left. Hence ${\cal P}$ consists of all the operators with $\Delta> \delta_{\rm crit}D$. We can further optimize over the kernel $f(z)$ to produce the lowest upper bound on the OPE coefficient squared $c^2_{\phi\phi {\cal O}_b}$. This optimization problem is the same one as is \cite{Mazac:2018mdx}. Borrowing their result,  
\begin{equation}\label{kernalfeg}
  f_{\rm opt}(z) = \frac{ (1-2z)}{\left[ z(z-1)\right]^{1/2} \left[ \sqrt{z(1-z)} + \sqrt{z_b(1-z_b)} \right]^2} \ , \qquad z_b= z_*(\delta_b).
\end{equation} 
For the case of stress tensor, we use $\delta_b=1$ and the equation \eqref{ct} that relates $c^2_{\phi\phi T^{\mu\nu}}$. 

\subsection*{Case II: $\delta_\phi>3/4$}
In this case, the kernel $f_{\rm opt}(z)$ leads to a functional that is negative for the stress tensor. However, we need the functional to be positive on the stress tensor to obtain lower bound on the central charge. This is achieved simply by using $f(z)=-f_{\rm opt}(z)$ as the kernal. With this choice, the functional is positive for $\delta<\delta_{\rm crit}$ and also for $\delta>2\delta_\phi$ but it is negative for $\delta_{\rm crit}<\delta<2\delta_\phi$. The schematic plot of this functional is shown in figure \ref{lessthangreaterthanthreequarters} on the right.
The region $(\delta_{\rm crit},2\delta_\phi)$ is excluded from the set ${\cal P}$. In other words, for the case $\delta_\phi>3/4$, the set ${\cal P}$ consists of operators with $\delta<\delta_{\rm crit}$ along with operators satisfying $\delta>2\delta_\phi$.

\subsection{Modifying the set ${\cal P}$}
We can modify the set ${\cal P}$ by changing the functional in both of these cases. This is important for widening the applicability of the bound i.e. to make it so that all the CFT operators ${\cal S}\subset {\cal P}$.
Let us call the function that satisfies all the properties listed below equation \eqref{1Dbounds} a kernel function. Note that if we multiply any kernel function by the so called CDD factor $\alpha(z,z_i)$ we get a new kernel function. The CDD factor is given by 
\begin{align}
  \alpha(z,z_i)=\tilde \alpha (x(z),x(z_i)), \qquad \tilde \alpha(x,y)=\frac{x-y}{xy-1}, \quad x(z)=\frac{\sqrt{z_b(1-z_b)}-\sqrt{z(1-z)}}{\sqrt{z_b(1-z_b)}+\sqrt{z(1-z)}}
\end{align}
where $z_i\in (0,1)$. It has the property that $|\alpha(z,z_i)|=1$ for $z\in (-\infty, 0)$.
The $\alpha(z,z_i)$ factor endows the kernel with a pair of additional single zeros in the unitarity region. These two new zeros $\delta_i$ are at the two solutions of the equation $z_i=z_*(\delta_i)$ and $1-z_i=z_*(\bar \delta_i)$, respectively. Each of these equations have two solutions but we are interested in the ones that lie in the unitarity domain.
 The pair of zeros $(\delta_i,\bar \delta_i)$ are related to each other as 
\begin{equation}\label{mirror}
  \bar{\delta_i} = \frac{1}{2} \left(1+\sqrt{1-4 (\delta_i -1) \delta_i +4 \delta \phi  (4 \delta \phi -3)}\right).
  \end{equation}
This follows easily from equation \eqref{saddlez}.
It is easy to see that when both the roots are real, they are on opposite sides of $\delta_{\rm crit}$. 
We can think of any one of the zeroes as the mirror image of the other across the point $\delta_{\rm crit}$.
This is because a point that is closer to $\delta_{\rm crit}$ on the right side is mirrored to a point that is closer to $\delta_{\rm crit}$ on the right. One can also see (by definition) that the mirror image of $\delta_{\rm crit}$ is $\delta_{\rm crit}$.  Also, the point $\delta=1/2$ is mirrored to the point $\delta= \bar \delta_{1/2} \equiv  \frac{1}{2} \left(1+\sqrt{2 +4 \delta \phi  (4 \delta \phi -3)}\right)$. Hence, this mirror maps the interval ${\cal D}_1 \equiv (0,\delta_{\rm crit})$ to the interval ${\cal D}_2 \equiv (\delta_{\rm crit}, \bar \delta_{1/2})$. Also for $\delta \in {\cal D}_3 \equiv (\bar \delta_{1/2}, 2 \delta_{\phi})$ the solution $\bar \delta$ is complex and does not correspond to any zero of the functional for real $\delta$.
\begin{figure}[t]
  \begin{center}
    \includegraphics[scale=0.5]{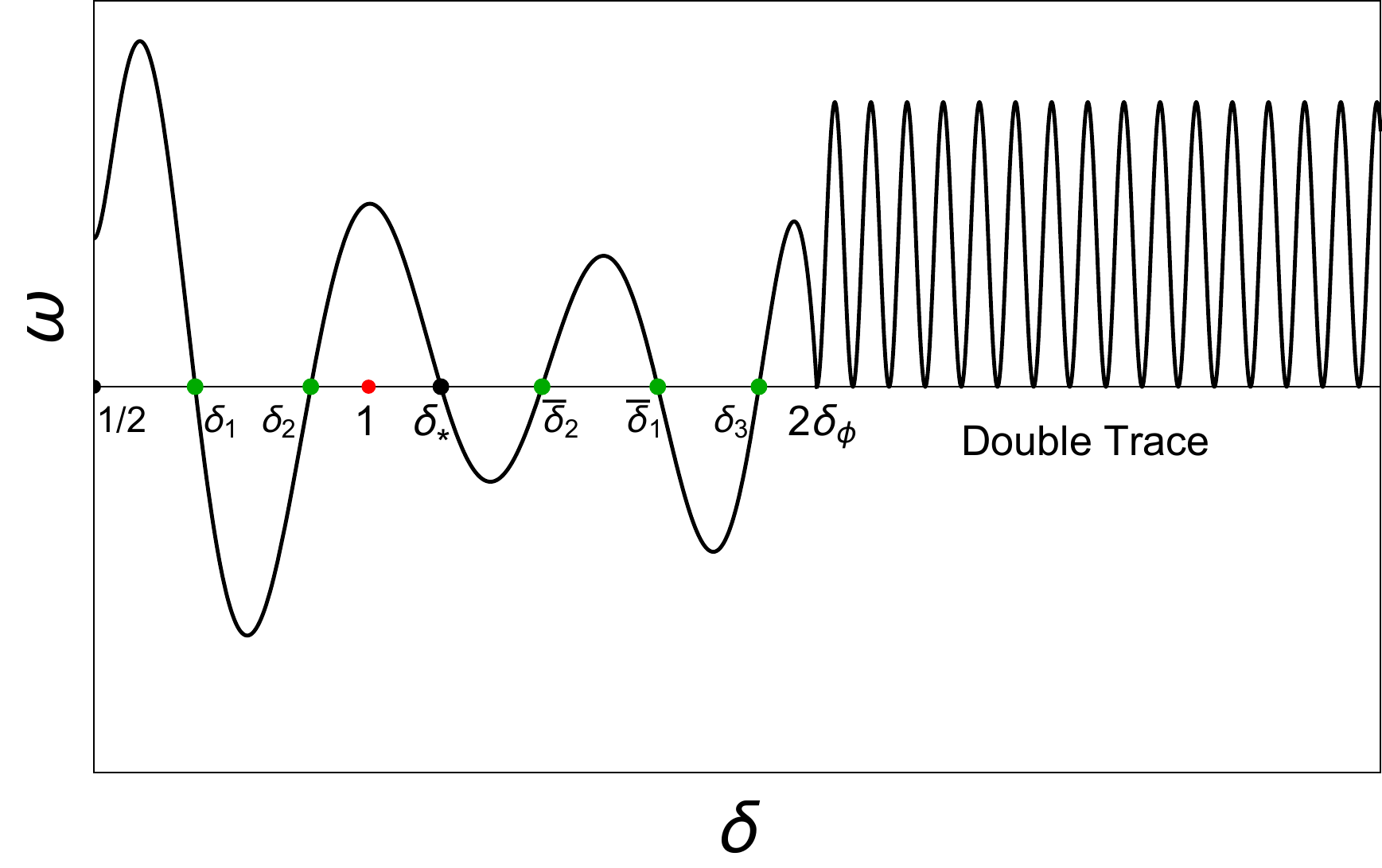}
  \end{center}
    \caption{Schematic plot of functional $\omega(\delta,\ell)$ 
    where the kernel function $f(z)$ is taken to be $f_{\rm opt}(z)$ times three CDD factors. There are two single zeros in ${\cal D}_1$ and two ``mirror image'' single zero in ${\cal D}_2$. The function as double zeros for $\delta>2\delta_\phi$. The unpaired zero $\delta_3$ is in region ${\cal D}_3$. The CDD factors are chosen in a way that the functional is positive for stress tensor i.e. $\delta=1$.}\label{Eg3figure} 
  \end{figure}
In figure \ref{Eg3figure} we have displayed a schematic plot of the functional obtained after multiplying $f_{\rm opt}(z)$ by three CDD factors. We can see that the set ${\cal P}$ now consists of points on either sides of  $\delta_{\rm crit}$. As we change the number of CDD factors and their parameters $z_i$ the set ${\cal P}$ changes. This gives us a lot of freedom to tune the ${\cal P}$ so that the set of CFT operators ${\cal S}\subset {\cal P}$. However, as we will see below, one can not do so in all the cases.
To this end, let us characterize the CFT spectrum ${\cal S}$ for which it is not possible, in our framework, to obtain a functional with ${\cal P}$ such that ${\cal S}\subset {\cal P}$.

\subsection{Applicability of the bound}\label{cddfactor}
First notice that even if we use the CDD factors to multiply the kernel \eqref{kernalfeg} the functional takes opposite values on the mirror pairs. This is simply a consequence of anti-symmetry $f(z)=-f(1-z)$. 

\begin{figure}[t]
  \begin{center}
    \includegraphics[scale=0.25]{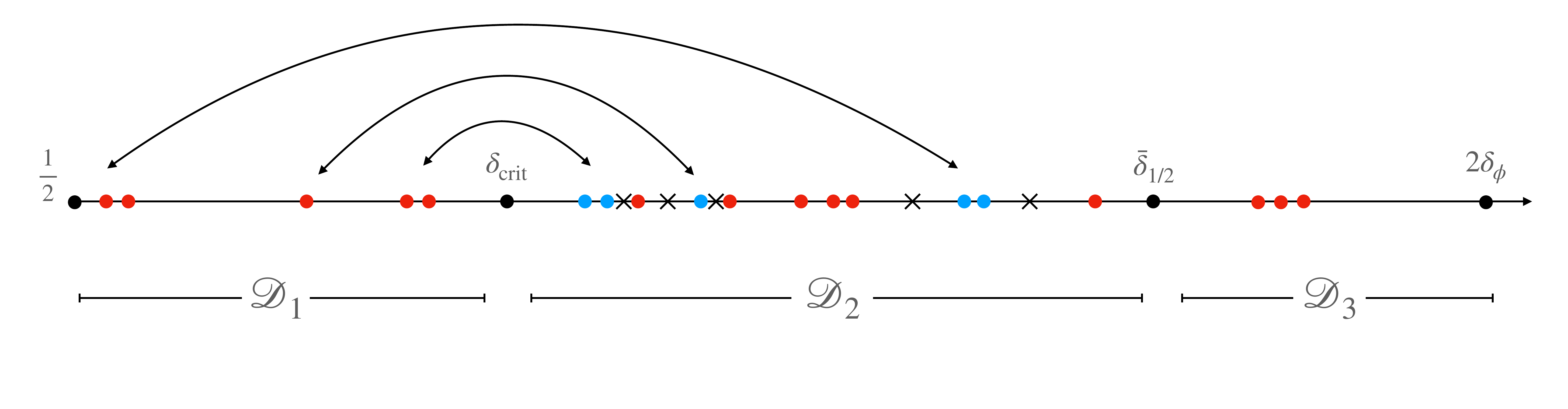}
  \end{center}
    \caption{The CFT operators are denoted on the $\delta$ axis with red dots. The blue dots are the ``mirror reflection'' of the red dots in the region ${\cal D}_1$ to the region ${\cal D}_2$. In order to construct a functional with ${\cal S}\in {\cal P}$, we pick five CDD factors with zeros in the red-to-blue transition regions. These are as shown in the figure schematically with crosses.}\label{red-blue} 
\end{figure}

For the rest of the discussion,  it is useful to think graphically. Let us denote all the CFT operators on the $\delta$ axis with red dots. Let us reflect all the red dots in the ${\cal D}_1$ region onto the ${\cal D}_2$ region. Let us color these reflections blue. Now focus only on the region ${\cal D}_2 \cup {\cal D}_3$. It has some distribution of red dots and blue dots, with only red dots in region ${\cal D}_3$. We will be successful in finding the desired ${\cal P}$ if the functional is positive on the red dots and negative on the blue dots. To achieve this we multiply the kernel \eqref{kernalfeg} by CDD factors that give a single zero whenever we transit from red dots to blue dots and vice versa so that the functional is positive on the red dots and negative on the blue ones. This construction is illustrated  in figure \ref{red-blue}.
In the unlikely case there are pairs of blue dot and red dot that are coincident we need to have a single zero precisely at that point. 
This functional achieves our objective of ${\cal S}\subset {\cal P}$ with the minimum number of CDD factors. When is this construction not admissible? If the total number of CDD factors used are ${\cal O}(D)$ then the saddle point computation performed in equation \eqref{saddlez} is invalid as it is no longer determined only by the conformal block. So we conclude that the exponential lower bound \eqref{Ctbound} on the central charge is applicable if the following conditions are met:
\begin{itemize}
  \item {\bf Condition 1:} The exponential lower bound \eqref{Ctbound} on the central charge is applicable if there are less than ${\cal O}(D)$ transitions between red dots and blue dots. 
  \item {\bf Condition 2:} There is no CFT operator that is in the ${\cal O}(1/D)$ neighborhood in the $\delta$ space of the mirror image of the stress tensor $(\delta=\bar \delta_T \equiv  \frac{1}{2} \left(1+\sqrt{1 +4 \delta \phi  (4 \delta \phi -3)}\right))$. 
\end{itemize}
The second condition, in particular implies that for $\delta_\phi=3/4$, $\bar \delta_T=\delta_T=1$ and hence the bound is not obtained.
Is the first condition reasonable? We can answer this question with the same level of (im)precision with which it is asked. The following discussion of this condition is somewhat impressionistic and is only meant to offer some intuition.

Let $\phi$ be the CFT operator with smallest scaling dimension $\delta_\phi D$.
For $\delta_\phi<3/4$, we have $\delta_{\rm crit} <1$. In this case, the operators in the region ${\cal D}_1$ i.e. between $ (1/2,\delta_{\rm crit})$ are only scalars because of the unitarity bound on operators with spin. Because they are only scalars, it is reasonable to assume that these are less than ${\cal O}(D)$ in number.
On the other side of $\delta_{\rm crit}$, beyond $\delta=1$, we have operators that have arbitrary spin. It is perhaps natural that there are ${\cal O}(D)$ of them in the regions ${\cal D}_2$ and $ {\cal D}_3$ i.e. between $(\delta_{\rm crit}, \bar \delta_{1/2})$ and $(\bar \delta_{1/2}, 2\delta_\phi)$ respectively. Because the applicability of the bound only depends on the number of red-blue transitions as described above and that the number of operators are expected to be less than ${\cal O}(D)$ operators in ${\cal D}_1$, we expect the bound on the central charge to be applicable.

For $\delta_\phi >3/4$, we have $\delta_{\rm crit} >1$. In this case, on both side of $\delta_{\rm crit}$ i.e. in all three regions ${\cal D}_1$, ${\cal D}_2$ and ${\cal D}_3$ there are spinning operators. Operators in both these regions could perhaps be ${\cal O}(D)$ in number. In this case, it is not reasonable expect that the number of red-blue transitions are less than ${\cal O}(D)$ in number. So in this case we have to assume, perhaps somewhat unnaturally, that the spectrum operators in the window $(1,\delta_{\rm crit})$ is sparse i.e. of less than ${\cal O}(D)$ for the central charge bound \eqref{Ctbound} to be valid. As $\delta_\phi$ increases above $3/4$, at $\delta_\phi\sim 1.44$ we get $\delta_{\rm crit}=2$. At this point, we definitely have multi stress-tensor operators appearing in the region ${\cal D}_1$ in a dense fashion. We don't expect the central charge bound to be valid in this case. However, we don't need to probe the regime $\delta_\phi>1$ at all. If the lightest scalar has $\delta_\phi >1$, we can simply use the stress tensor component, say $T^{00}$ as a scalar operator in $D-1$ dimensional subspace. As $D$ is taken to be large, this change in dimension is of little significance. For $\delta_{\phi=T^{00}}=1$, $\delta_{\rm crit}\sim1.37$. For the bound to be applicable we have to require that the operator spectrum should be sparse up to this value of $\delta_{\rm crit}$.

This discussion also makes it clear why the free theory is allowed in large $D$ from this point of view. If we consider the four point function of the  free field $\phi$, we have $1=2\delta_{\phi}$. At this value of $\delta$, the functional vanishes and we do not get any bound  on the  central charge. We could also consider the four point function of $\phi^2$ operators. In this case, $\delta_{\phi^2}=1$, this means $\delta_{\rm crit}=1.37$. However, it is evident that a dense double-trace spectrum starts from $\delta=1$ and hence the sparseness assumption and in turn the central charge bound is not applicable.

In \cite{Gadde:2020nwg} a different approach was taken to constrain the space of conformal theories in large dimension. The conclusions from that paper agree with the results presented in this paper. In \cite{Gadde:2020nwg} it was argued that for $\delta_\phi<3/4$, the four point function of $\phi$'s is exponentially close to that of the generalized free field theory. For $3/4<\delta_\phi<1$, this conclusion required a sparseness assumption on the spectrum akin to the one argued here.

\section{Corrections at large but finite $D$}\label{sectionNumericalBounds}
Now that we have obtained exponential bound in large $D$ at leading order, it is a natural to ask how the bound gets corrected at finite $D$. 
This is essentially a question about correction to the functional value $\omega(\Delta,\ell)$.
There are two sources of corrections. 
\begin{itemize}
  \item Corrections to the kernel $h(z)$.
  \item Corrections to the integral \eqref{1dfuncG} with the corrected kernel.
\end{itemize}
Let us discuss the first source of corrections. We would like the kernel to obey the gluing condition \eqref{gluingcond} and the extremality condition. Note that the  condition that the inequality in euqation \eqref{extremalcond} is saturated is not a true extremal condition in finite dimension since it 
does not give rise to double zeros of $\omega(\Delta,\ell)$ for $\Delta\geq 2\Delta_\phi$. It does however insure that the functional is non-negative for $\Delta\geq 2\Delta_\phi$. It is in principle possible to solve the true extremality condition in perturbation theory in $1/D$ but it is quite cumbersome to do so. We will instead take the equation \eqref{extremalcond} as a necessary condition on the kernel in finite dimension as it at least guarantees positivity of $\omega(\Delta,\ell)$ for $\Delta\geq 2\Delta_\phi$. The gluing condition and ``extremality condition'' together imply
\begin{align}\label{gluingrel}
  {\rm Im}(f(z))=z^{2D\delta_\phi -2}|f(1/z)|+(1-z)^{2D\delta_\phi -2}|f(1/(1-z))|\qquad \qquad z\in (0,1).
\end{align}
In our analysis we could simply set the right hand side to zero in the $D\to \infty$ limit. In large but finite $D$, the right hand side does give small non-perturbative corrections. We will ignore these corrections and continue to use the kernel $f_{\rm opt}(z)$ that satisfies ${\rm Im} f(z)=0$ for  $z\in (0,1)$ even in finite dimension. 

\begin{figure}[H]
  \begin{subfigure}{.5\textwidth}
    \hspace{2.5cm}
  \includegraphics[scale=0.22]{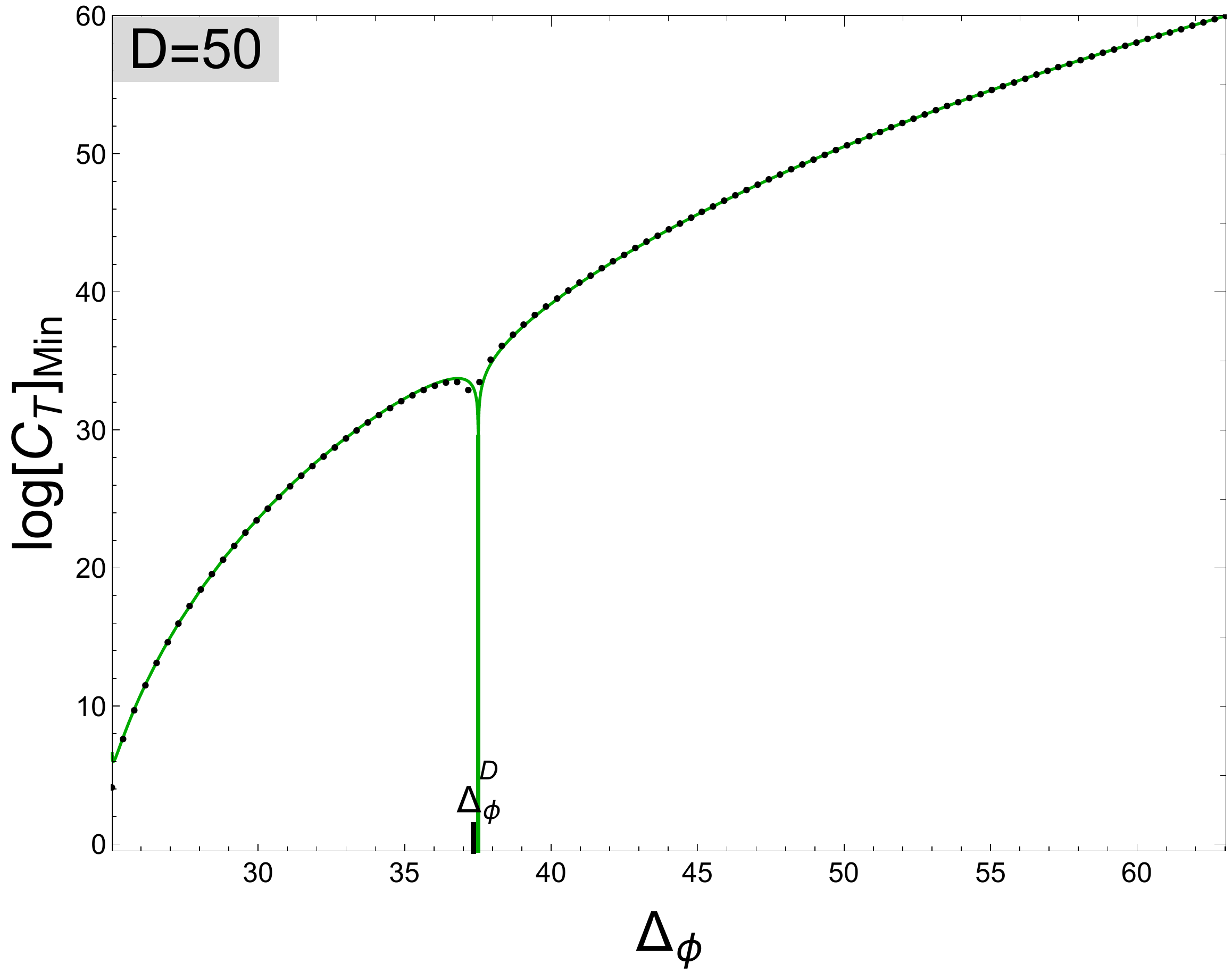}
  \end{subfigure} 
  \hspace{0.5cm}
  \begin{subfigure}{.5\textwidth}
  \includegraphics[scale=0.22]{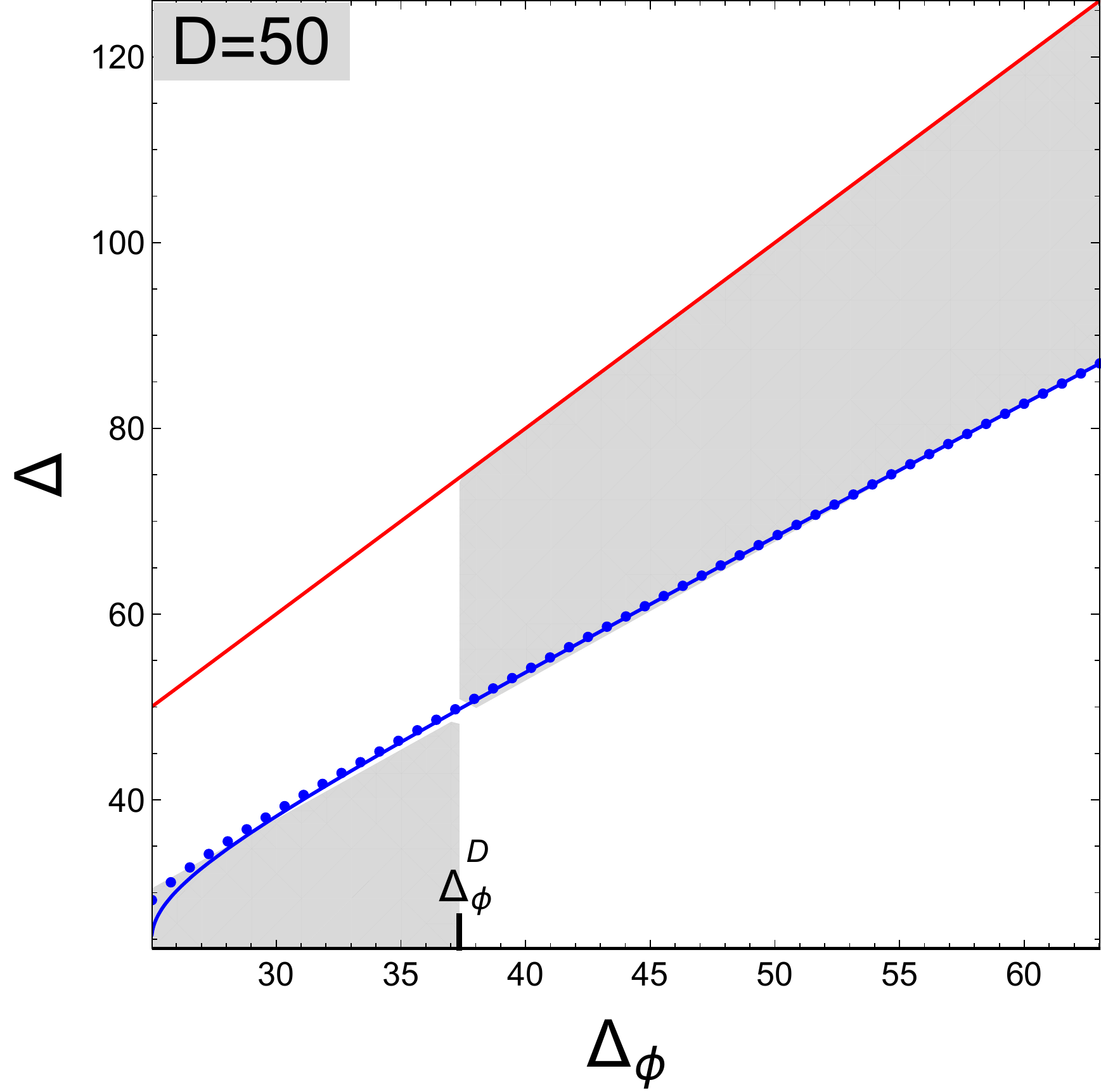}
  \end{subfigure}
  \\ \\
  \begin{subfigure}{.5\textwidth}
    \hspace{2.5cm}
  \includegraphics[scale=0.22]{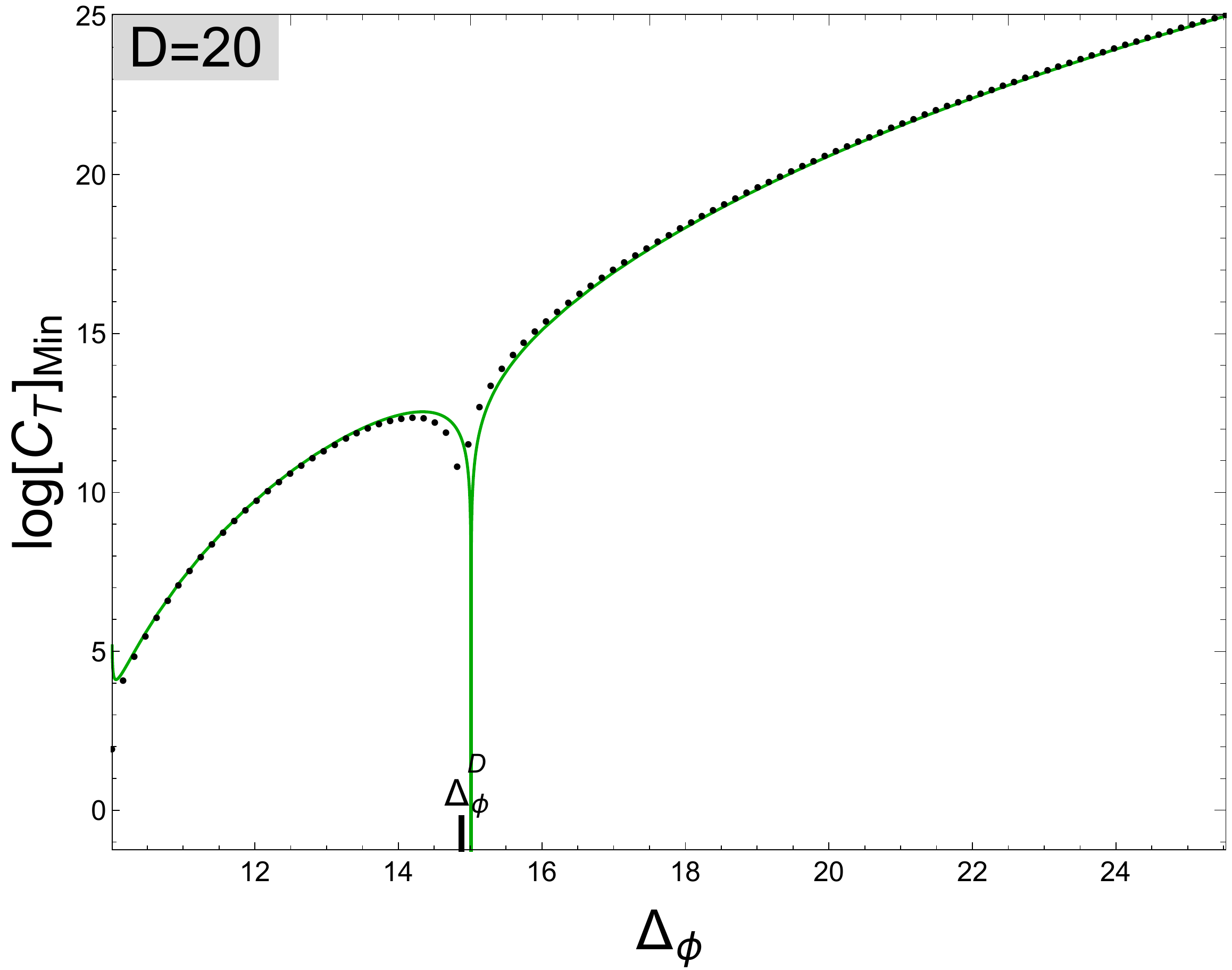}
  \end{subfigure} 
  \hspace{0.5cm}
  \begin{subfigure}{.5\textwidth}
    \includegraphics[scale=0.22]{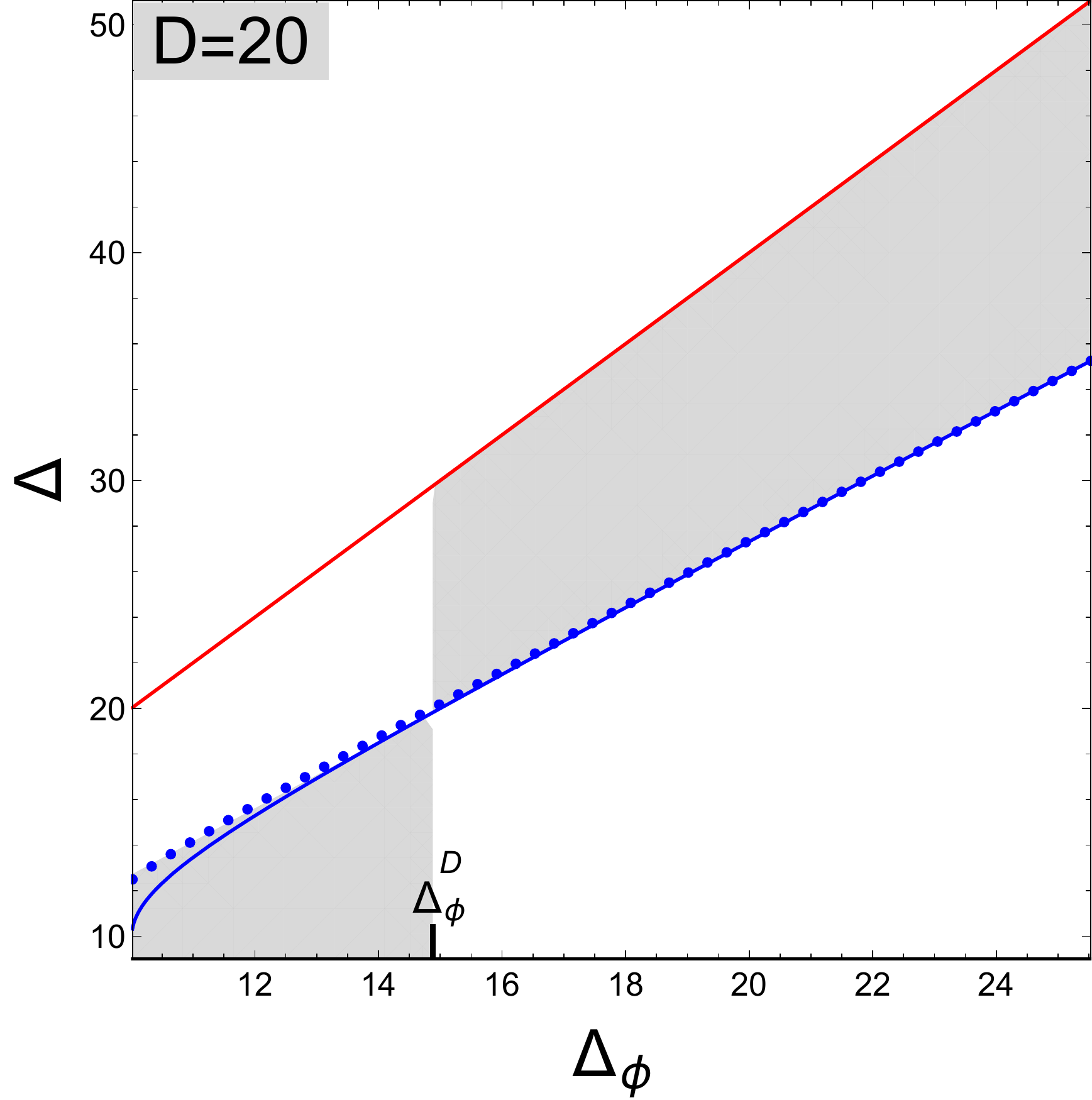}
    \end{subfigure}
  \\ \\
  \begin{subfigure}{.5\textwidth}
    \hspace{2.5cm}
  \includegraphics[scale=0.22]{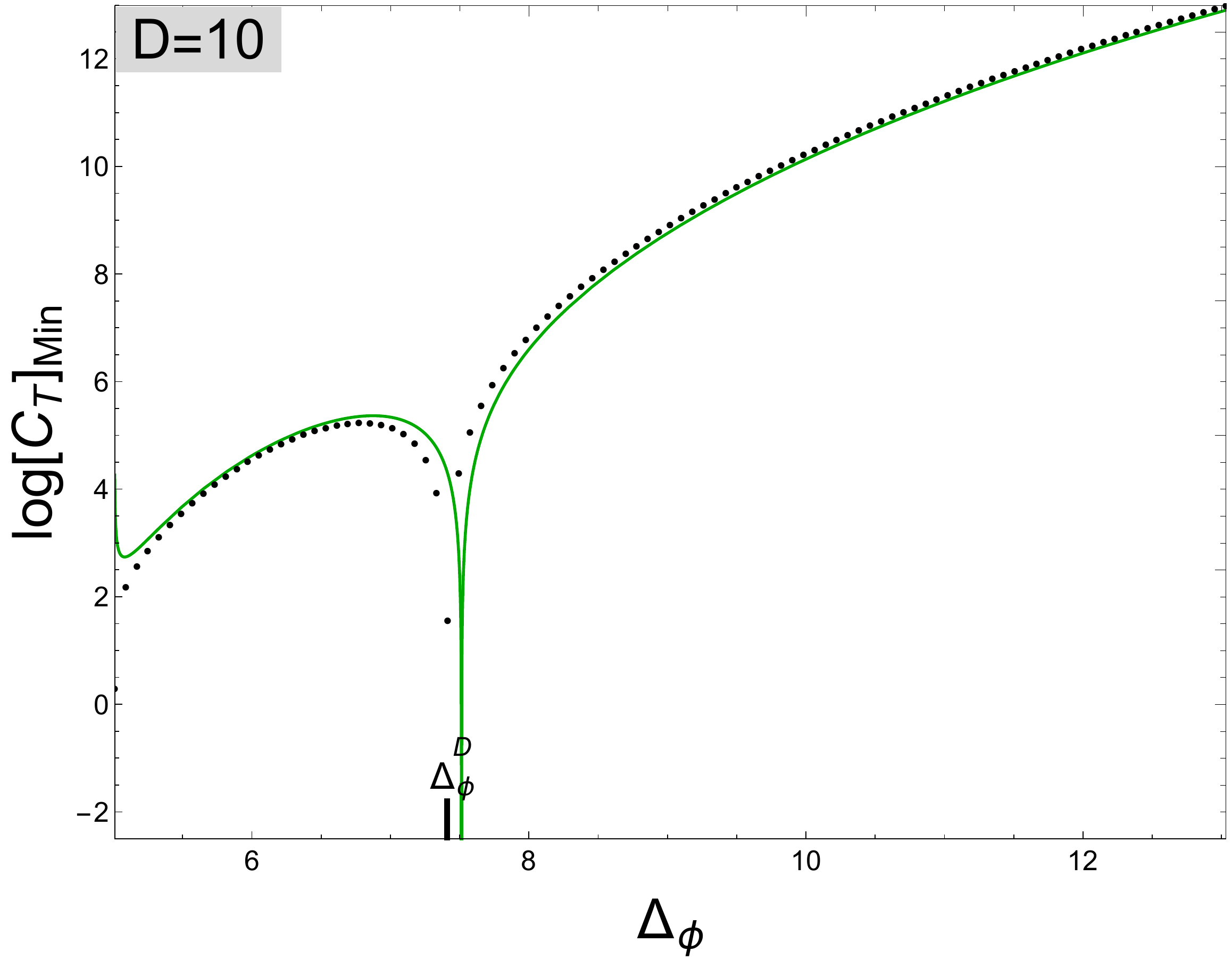}
  \end{subfigure} 
  \hspace{0.5cm}
  \begin{subfigure}{.5\textwidth}
    \includegraphics[scale=0.22]{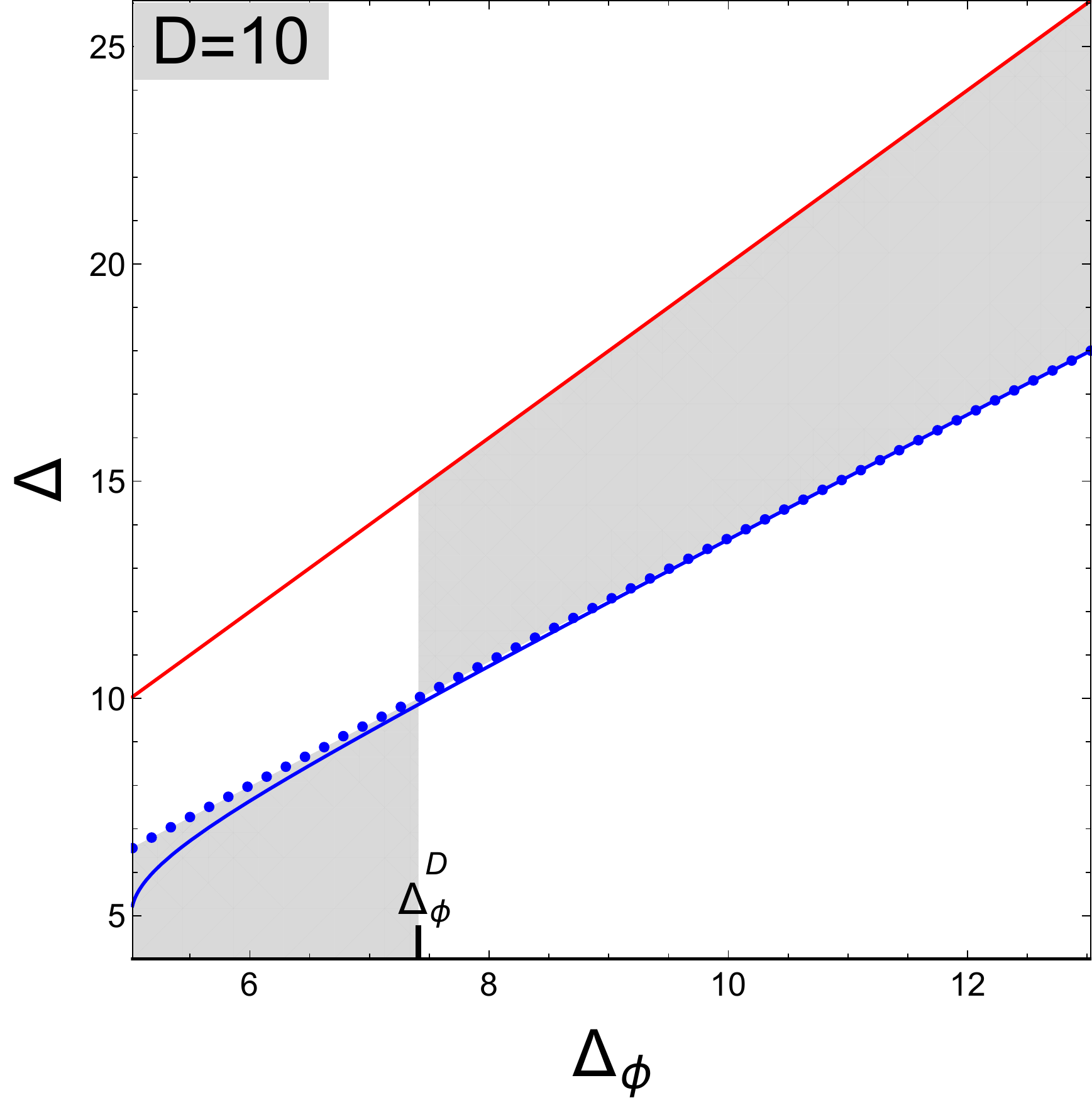}
    \end{subfigure}
  \caption{Approximate numerical bounds on the central charge of CFTs in $D=50, 20, 10$. The green curve is obtained by substituting appropriate value of $D$ in equation \eqref{Ctbound}. The bounds are obtained with the assumption that no scalar operators are present in the gray shaded range for a given $\Delta_\phi$ (similar but less constraining assumptions are also made for spinning operators). The red line is $\Delta=2\Delta_\phi$ and the solid blue line is $\Delta=D\delta_{\rm crit}$.} 
  \label{502010D}
  \end{figure}

\begin{figure}[H]
    \begin{subfigure}{.5\textwidth}
      \hspace{2.5cm}
    \includegraphics[scale=0.22]{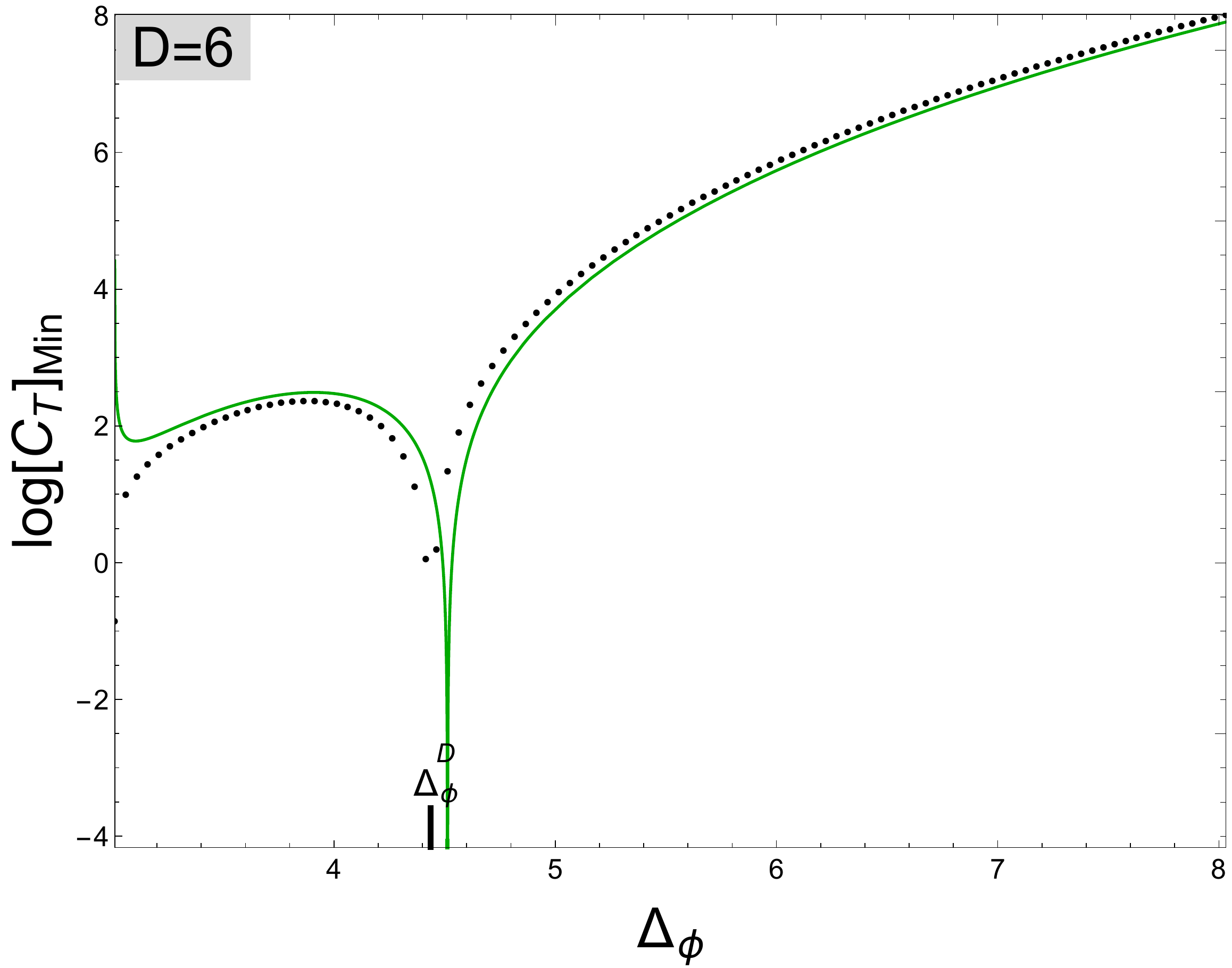}
    \end{subfigure}
    \hspace{0.5cm}
    \begin{subfigure}{.5\textwidth}
      \includegraphics[scale=0.22]{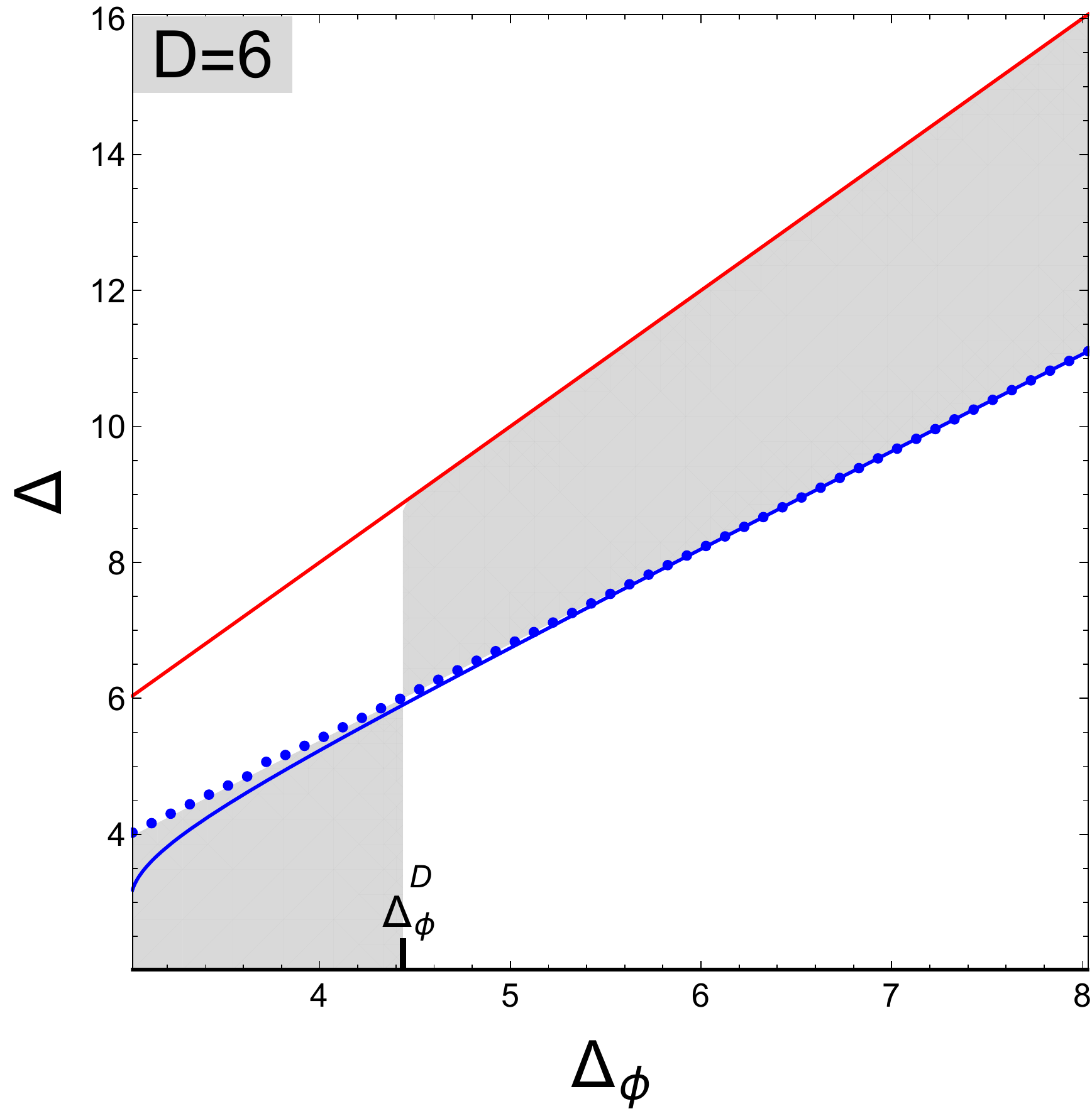}
      \end{subfigure}
    \\ \\
    \begin{subfigure}{.5\textwidth}
      \hspace{2.5cm}
    \includegraphics[scale=0.22]{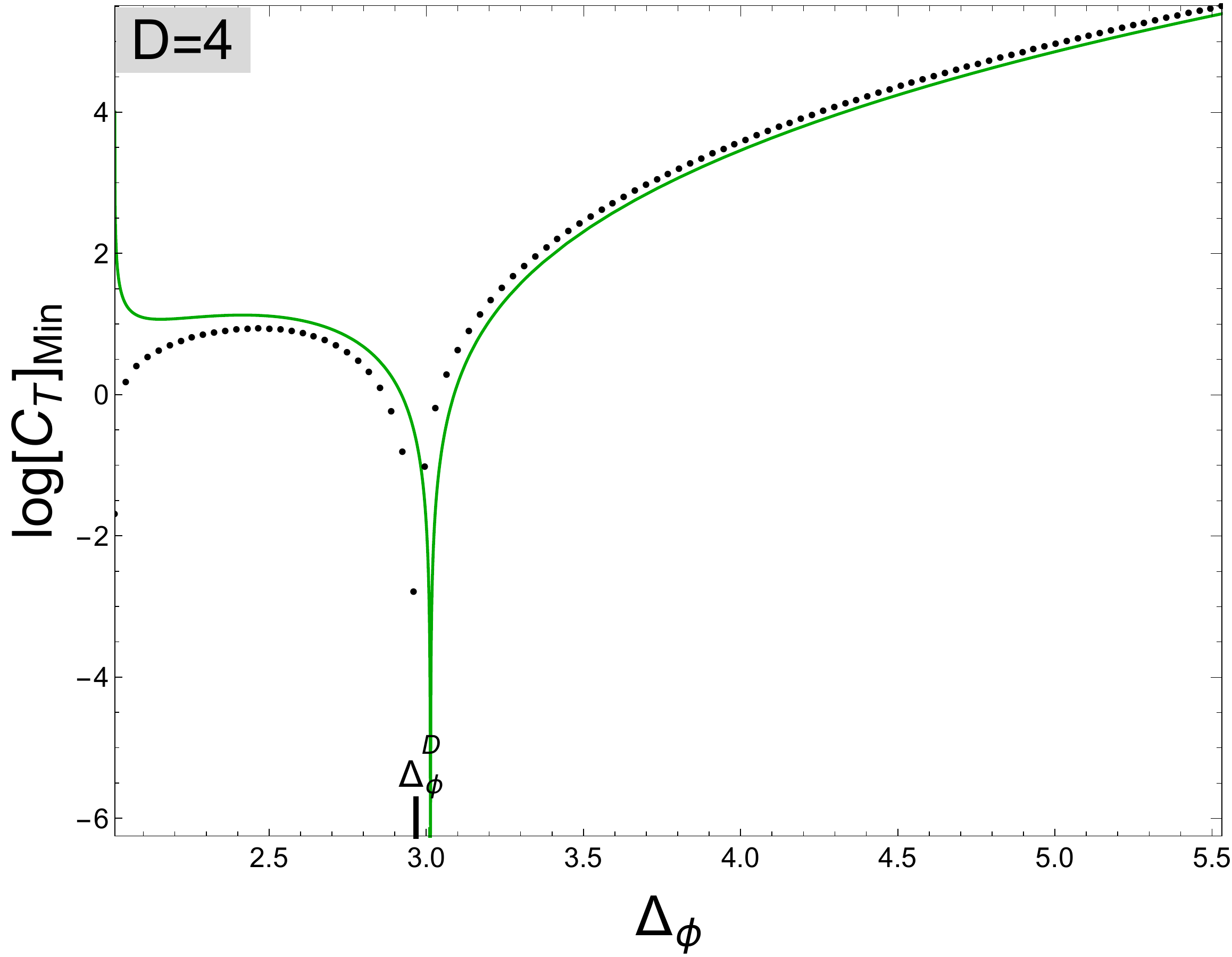}
    \end{subfigure}
    \hspace{0.5cm}
    \begin{subfigure}{.5\textwidth}
      \includegraphics[scale=0.22]{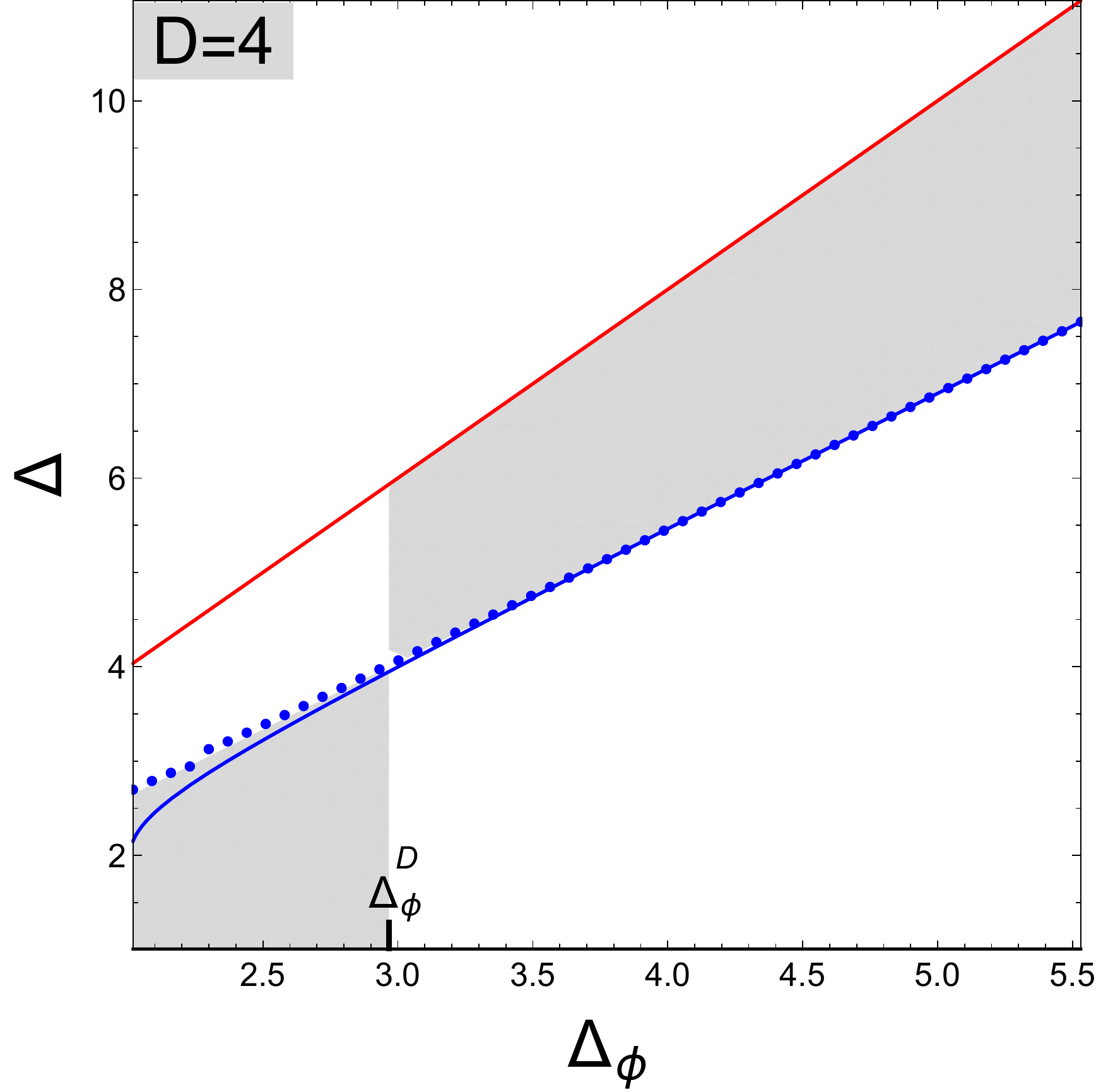}
      \end{subfigure}
    \caption{Approximate numerical bounds on the central charge of CFTs in $D=6, 4$. The green curve is obtained by substituting appropriate value of $D$ in equation \eqref{Ctbound}. The bounds are obtained with the assumption that no scalar operators are present in the gray shaded range for a given $\Delta_\phi$ (similar but less constraining assumptions are also made for spinning operators). The red line is $\Delta=2\Delta_\phi$ and the solid blue line is $\Delta=D\delta_{\rm crit}$.} 
    \label{64D}
    \end{figure}

The second set of corrections, viz. the corrections to the saddle point integral \eqref{1dfuncG} are both perturbative as well as non-perturbative. Instead of treating them differently, we deal with them by simply performing the integral numerically. We expect to get close to accurate bounds (with errors that are exponentially small in $D$) in large but finite dimension in this way. The results for $D=50,20,10,6,4$ are presented in figure \ref{502010D}, \ref{64D}. We have also given the bound obtained by setting the appropriate value of $D$ in equation \eqref{Ctbound} with $f(z)=f_{\rm opt}(z)$ for reference. As we have not made use of the CDD factors in the functional, we have to assume absence of operators in certain $\Delta$ range. This range is given by gray shaded region in the figures. In principle, we can tune the functional to the operator spectrum using the CDD factors as described  in section \ref{cddfactor}.

The only reason that these bounds are not completely trustworthy for small $D$ is because the right hand side of equation \eqref{gluingrel} can not be approximated by zero for small $D$. 
Also note that $D$ controls these errors only because we have take $\Delta_\phi$ to be of the same order as $D$. This implies that the numerical bounds are good approximations even in small $D$ as long as $\Delta_\phi$ is taken to be large. In this way we can repurpose these bounds as bounds on the central charge for theories with large gap. We are currently investigating this direction. In order to get completely trustworthy bounds for small $D$ and small $\Delta_\phi$,  we need to solve the condition \eqref{gluingrel} exactly.

\section*{Acknowledgement}

We would like to thank the TIFR string theory group, in particular Gautam Mandal, Shiraz Minwalla, Onkar Parrikar, Sandip Trivedi for useful discussions. We would like to thank Adwait Gaikwad for collaboration in related projects. This work is supported by the Infosys Endowment for the study of the Quantum Structure of Spacetime. The work of A.G. is also supported by the SERB Ramanujan fellowship. We acknowledge support of the Department of Atomic Energy, Government of India, under Project Identification No. RTI 4002. We would also like to acknowledge our debt to the people of India for their steady support to the study of the basic sciences.

\appendix

\section{Simplifying the functional}\label{functional-simple}

In this appendix, we will simplify the functional given in equation \eqref{1dfuncG} to the expression in \eqref{simplifiedw} using some simple properties of the conformal block specialized to the locus $z=\bar z$. We denote the block specialized to this locus as $G_{\Delta,\ell}(z)$.
\begin{align} \label{1dblkprop}
    G_{\Delta,\ell}(z\pm i \epsilon) &= e^{\pm i \pi \Delta} \hat{G}_{\Delta, \ell}(z) \ ,  \qquad \qquad \ z \in (-\infty,0) \notag \\
    G_{\Delta, \ell}(z) &= \hat{G}_{\Delta, \ell} \left( \frac{z}{z-1} \right) \ , \qquad \qquad z \in (0,1) \qquad \textrm{(u-symmetry)}  . 
\end{align}
Where $\hat{G}_{\Delta, \ell} = |G_{\Delta, \ell}(z+ i \epsilon)|$, for $z \in (-\infty,0) $. Using these properties it is easy to see that the first term in \eqref{1dfuncG} simplifies to,
\begin{equation}\label{1dsimpt1}
\int_{-\infty}^{0} d z \hspace{0.15cm} h(1-z) \textrm{Disc}\left[ \frac{G_{\Delta, \ell}(z)}{z^{2\Delta_{\phi}}}  \right] = \left( e^{i \pi (\Delta - 2 \Delta_{\phi})}-e^{-i \pi (\Delta - 2 \Delta_{\phi})}  \right) \int_{-\infty}^{0} d z \hspace{0.15cm} h(1-z) \frac{\hat{G}_{\Delta, \ell}(z)}{|z|^{2\Delta_{\phi}}}.
\end{equation} 
To simplify the second term we first write the integral over the discontinuity as a contour integral wrapping the branch cut from $(-\infty,0)$. Then we do a variable change $z\rightarrow 1-z$ and deform the new integration contour that wraps the cut $(1,\infty)$ back to the one that wraps the cut $(-\infty,0)$. In doing so we  use the fall off condition of $h(z)$ given above equation \eqref{1dfuncG} to neglect the contribution of the arc at infinity. In this deformation process, we also collect the contribution of the branch cut of $h(z)$ in $(0,1)$. 
\begin{align}\label{1dsimpt2}
 & \int_{-\infty}^{0} d z \hspace{0.25cm} h(1-z) \textrm{Disc}\left[ \frac{G_{\Delta, \ell}(1-z)}{(1-z)^{2\Delta_{\phi}}} \right] \\ 
     =&  \int_{-\infty}^{0} d z \hspace{0.25cm}  \textrm{Disc}\left[h(z) \frac{G_{\Delta, \ell}(z)}{z^{2\Delta_{\phi}}} \right] +   \int_{0}^{1} d z \hspace{0.25cm}  \textrm{Disc}\left[h(z)  \right] \frac{G_{\Delta, \ell}(z)}{z^{2\Delta_{\phi}}} \endline
=&  \int_{-\infty}^{0} d z \left(h(z+i \epsilon) e^{i \pi (\Delta - 2 \Delta_{\phi})} - c.c. \right)  \frac{\hat{G}_{\Delta, \ell}(z)}{|z|^{2\Delta_{\phi}}} + \int_{0}^{1} d z \hspace{0.15cm}   \textrm{Disc}\left[h(z)  \right] \frac{G_{\Delta, \ell}(z)}{z^{2\Delta_{\phi}}} \notag \\
=& \int_{-\infty}^{0} d z \left(h(z+i \epsilon) e^{i \pi (\Delta - 2 \Delta_{\phi})} - c.c. \right)  \frac{\hat{G}_{\Delta, \ell}(z)}{|z|^{2\Delta_{\phi}}} \notag\\
+& \int_{-\infty}^{0}d z \hspace{0.25cm} (1-z)^{2\Delta_{\phi}-2} \textrm{Disc}\Big[ h\left(\frac{z}{z-1}\right) \Big] \frac{\hat{G}_{\Delta, \ell}(z)}{|z|^{2\Delta_{\phi}}} \notag
\end{align}
In second term of last line we have done variable transform from $z \rightarrow z/(z-1)$ and used the u-symmetry of block from equation \eqref{1dblkprop}. Since $h(z)$ is real for $z\in (1,\infty)$ we have $h(z^{\ast})= h(z)^{\ast}$. This implies $\textrm{Disc}\left[h(z)  \right]$ is purely imaginary. Combining \eqref{1dsimpt1} and \eqref{1dsimpt2}, and using the definitions of $f(z)$ and $g(z)$ in equations \eqref{1df} and \eqref{1dg} respectively, gives the simplified expression in equation \eqref{simplifiedw}.

\subsection*{Action on operators with $\Delta>2\Delta_{\phi}$ }
The simplified integral \eqref{simplifiedw} can be sued to compute the functional for operators with $\Delta \geq 2\Delta_\phi$. That is because both ${\mathfrak f}$ and ${\mathfrak g}$ terms are separately convergent in this range. In the large $D$ limit, these integrals are performed using the saddle point approximation. As $\tilde g(z)\equiv (1-z)^{2\Delta_\phi} g(z)$ is take to be ${\cal O}(1)$, the saddle point is controlled by  $\hat{G}_{\Delta}(z)/|z|^{2 \Delta_{\phi}}$. This is computed in appendix \ref{blocksaddlepoint} in equation \eqref{saddleD}. The saddle point $z_*\leq 0$ for all $\Delta\geq 2\Delta_\phi$.

\subsection*{Action on operators with $\Delta<2\Delta_{\phi}$ }

\begin{figure}[t]
  \begin{subfigure}{.5\textwidth}
    \centering
    \includegraphics[scale=0.4]{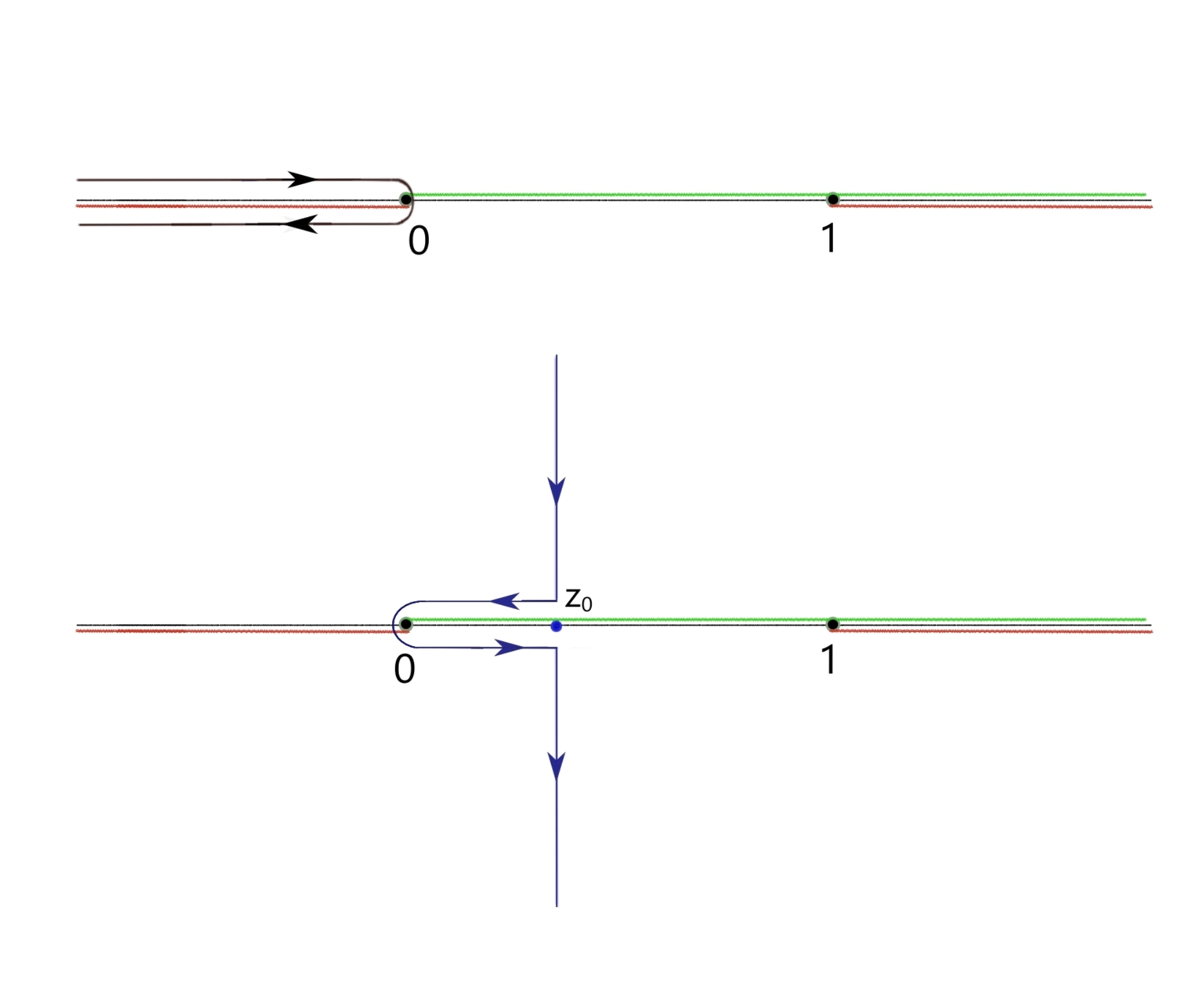}
  \end{subfigure}%
  \begin{subfigure}{.5\textwidth}
    \centering
  \includegraphics[scale=0.4]{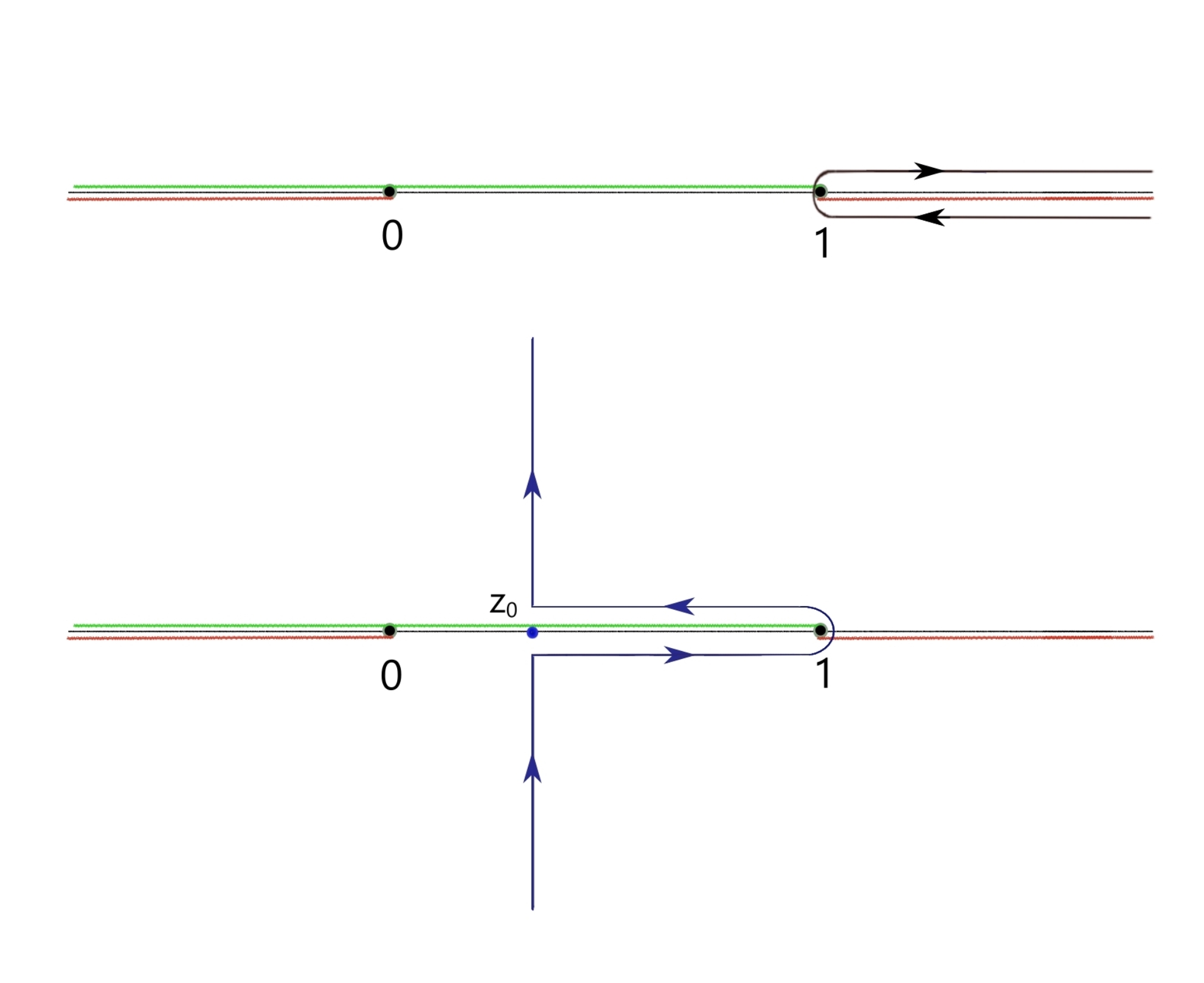}
  \end{subfigure}
  \caption{The top two contours are the original contours of integration in equation \eqref{funcb2}. The bottom two contours are their respective deformations to saddle points. } 
  \label{contourdeform}
  \end{figure}

In this regime, the two integrals of equation \eqref{simplifiedw} are not individually convergent. We need to resort to the original expression \eqref{1dfuncG} to compute the integral. 
\begin{align}\label{funcb2}
\omega(\Delta) &= \frac{1}{2 \pi i}\int_{-\infty}^{0} dz \ h(1-z)  \ \textrm{Disc} \left[\frac{G_{\Delta}(z)}{z^{2\Delta_{\Phi}}}- \frac{G_{\Delta }(1-z)}{(1-z)^{2\Delta_{\Phi}}} \right] 
\end{align}
If we look at the saddle point \eqref{saddleD}, it is easy to see that it lies in the range $(0,1)$ for $\Delta< 2\Delta_\phi$. This motivates the contour deformation shown in figure \ref{contourdeform}. 
The functional is written as,
\begin{align}
\omega(\Delta ) = & - \frac{1}{2 \pi i} \int_{0}^{z_{0}}dz \ \textrm{Disc}(h(1-z)) \ \frac{G_{\Delta }(z)}{z^{2\Delta_{\Phi}}} \ + \ \frac{1}{2 \pi i} \int^{1}_{z_{0}}dz \ \textrm{Disc}(h(z)) \ \frac{G_{\Delta }(z)}{z^{2\Delta_{\Phi}}} \notag \\
& + \frac{1}{2 \pi i} \int_{z_{0}}^{i \infty} dz \ (h(z)-h(1-z)) \ \frac{G_{\Delta }(z)}{z^{2\Delta_{\Phi}}} + \frac{1}{2 \pi i} \int^{z_{0}}_{-i \infty} dz \ (h(z)-h(1-z)) \ \frac{G_{\Delta }(z)}{z^{2\Delta_{\Phi}}} \notag \\
= & - \int_{0}^{z_{0}}dz \ g(1-z) \ \frac{G_{\Delta }(z)}{z^{2\Delta_{\Phi}}} \ - \ \int^{1}_{z_{0}}dz \ g(z) \ \frac{G_{\Delta }(z)}{z^{2\Delta_{\Phi}}} \notag \\
& + \frac{1}{2i} \int_{z_{0}}^{i \infty} dz \ f(z) \ \frac{G_{\Delta }(z)}{z^{2\Delta_{\Phi}}} \ + \ \frac{1}{2i} \int^{z_{0}}_{-i \infty} dz \ f(z) \ \frac{G_{\Delta }(z)}{z^{2\Delta_{\Phi}}} 
\end{align}
Here, in the second equality we have used the definitions of kernels $f(z)$ \eqref{1df} and $g(z)$ \eqref{1dg}. Using the scaling \eqref{fgscaling}, the first two terms in second equality (i.e. the $g$ terms) vanishes in the limit $\Delta_{\phi} \rightarrow \infty$. We can perform the saddle point integral in the direction perpendicular to real line at $z=z_{*}$ giving our integral,
\begin{equation}\label{funcinter}
\omega(\Delta ) =  \frac{\left( f(z_{*} + i \epsilon) +  f(z_{*} - i \epsilon) \right)}{2} \ \frac{G_{\Delta }(z_*)}{z_*^{2\Delta_{\Phi}}} \sqrt{\frac{2 \pi}{\Delta_{\phi} {\mathsf q}_\delta''(z_*)}}.
\end{equation}
In appendix \ref{blocksaddlepoint} we show that the steepest descent at $z_*$ is indeed perpendicular to the real axis.
Here the function ${\mathsf q}_\delta$ is defined in equation \eqref{qv}. Using the gluing condition \eqref{gluingcond} Im$(f(z))=0$, for $z\in(0,1)$. Thus, the functional in \eqref{funcinter} becomes
\begin{equation}
\omega(\Delta ) =  \textrm{Re}\left( f(z_{*}) \right) \ \frac{G_{\Delta }(z_*)}{z_*^{2\Delta_{\Phi}}} \sqrt{\frac{2 \pi}{\Delta_{\phi} p''(z_*)}}. 
\end{equation}

\subsection*{Action on Identity}
Here we will look at the action of functional \eqref{1dfuncG} on identity block given by,
\begin{equation}
\omega({\bf 1})=\frac{1}{2 \pi i } \int_{1}^{\infty} d z \hspace{0.25cm} h(z) \textrm{Disc} \left[ \frac{1}{z^{2\Delta_{\phi}}} - \frac{1}{(1-z)^{2\Delta_{\phi}}} \right] 
\end{equation}
The first term is zero since $z^{-2 \Delta_{\phi}}$ does not have any discontinuity in the region $(1,\infty)$. In the second term, we deform the contour from $(1,\infty)$ to $(-\infty,1)$, where $h$ has a discontinuity but $ (1-z)^{-2\Delta_{\phi}}$ does not.
\begin{align}
\omega({\bf 1}) &= \frac{1}{2 \pi i} \int_{-\infty}^{1} d z \frac{\textrm{Disc}\left[ h(z) \right]}{(1-z)^{2 \Delta_{\phi}}}   \endline
&= \frac{1}{2 \pi i} \int_{-\infty}^{0} d z \frac{\textrm{Disc}\left[ h(z) \right]}{(1-z)^{2 \Delta_{\phi}}}  + \frac{1}{2 \pi i} \int_{0}^{1} d z  \frac{\textrm{Disc}\left[ h(z) \right]}{(1-z)^{2 \Delta_{\phi}}} \endline
& = \frac{1}{2 \pi i} \int_{-\infty}^{0} d z \frac{\textrm{Disc}\left[ h(z) -h(1-z) \right]}{(1-z)^{2 \Delta_{\phi}}}  + \frac{1}{2 \pi i} \int_{0}^{1} d z  \frac{\textrm{Disc}\left[ h(z) \right]}{(1-z)^{2 \Delta_{\phi}}} \endline
& =- \int_{-\infty}^{0} d z \frac{\mathrm{Im}(f(z))}{(1-z)^{2 \Delta_{\phi}}}  - \int_{0}^{1} d z  \frac{g(z)}{(1-z)^{2 \Delta_{\phi}}} 
\end{align} 
where, in the third equality we have added $h(1-z)$ inside the discontinuity because $h(1-z)$ does not have any discontinuity on $z\in(-\infty,0)$. In the fourth equality we have used the definitions of kernels $f(z)$ \eqref{1df} and $g(z)$ \eqref{1dg}. Using the scaling \eqref{fgscaling}, the first term vanishes in the limit $\Delta_{\phi} \rightarrow \infty$, and we get 
 \begin{equation}
 \omega({\bf 1}) =   - \int_{0}^{1} d z  \frac{g(z)}{(1-z)^{2 \Delta_{\phi}}} = - \int_{0}^{1} d z \ \tilde{g}(z).
 \end{equation}
Now using extremality condition where the inequality \eqref{extremalcond} is saturated the above functional becomes,
\begin{equation}
 \omega({\bf 1}) =  - \int_{0}^{1} d z \ (1-z)^{-2} \left| f \left( \frac{z}{z-1} \right) \right| = - \int_{-\infty}^{0} d z \ |f(z)|.
\end{equation}

\section{Conformal blocks in large $D$ and saddle points}\label{blocksaddlepoint}

The conformal block in large dimension is computed in \cite{Fitzpatrick:2013sya}. On the  $z=\bar{z}$ locus it is given by 
\begin{equation}\label{Ldcb}
G_{\Delta,\ell}(z) = \frac{2^{\Delta }}{\sqrt{1-\left( \frac{z}{2-z}\right)^{2}}}A_{1-\ell}(1) \  \left( \frac{z}{2-z}\right)^{\Delta} \prescript{}{2}{F}_{1}\left(  \frac{\Delta}{2}, \frac{\Delta-1}{2}, \Delta-\frac{D}{2}+1, \left( \frac{z}{2-z}\right)^{2}  \right)
\end{equation}
where, the function $A_{\beta}(x)$ is given in equation \eqref{Afunction}. Using the integral representation of the hypergeometric function 
\begin{equation}\label{eulerint}
  B(b,c-b) \prescript{}{2}{F}_{1}\left( a,b,c,z\right) = \int_{0}^{1} dt \ t^{b-1}(1-t)^{c-b-1}(1-z t)^{-a} 
  \end{equation}
and the scaling $\Delta=\delta D$, the conformal block \eqref{Ldcb} becomes,
\begin{equation}
G_{\delta D,\ell}(z) = \frac{2^{\delta D }}{\sqrt{1-\left( \frac{z}{2-z}\right)^{2}}} A_{1-\ell}(1) \ \frac{ \left( \frac{z}{2-z}\right)^{\delta D}}{B \left( \frac{\delta D-1}{2}, \frac{\delta D-D+3}{2} \right)} \int_{0}^{1}  dt \ \sqrt{\frac{1-t}{t^3}} \left( \frac{t^{\frac{\delta}{2}}(1-t)^{\frac{\delta-1}{2}}}{\left( 1-\left( \frac{z}{2-z}\right)^{2}   t  \right)^{\frac{\delta}{2}}}  \right)^{D} 
\end{equation}
This integral can be done by saddle point approximation in large $D$ limit. This has been done in \cite{Gadde:2020nwg}, but we will reproduce it here  for convenience. The saddle point equation and its solution is,
\begin{equation}
\frac{d}{dt} \log \left( \frac{t^{\frac{\delta}{2}}(1-t)^{\frac{\delta-1}{2}}}{\left( 1-\left( \frac{z}{2-z}\right)^{2}   t  \right)^{\frac{\delta}{2}}}  \right) =0 \quad \Rightarrow \quad t_*^{\pm}= \frac{2 \delta-1 \pm \sqrt{4 \delta(\delta-1) \left(1-\left( \frac{z}{2-z}\right)^{2}  \right)+1}}{2(\delta-1)\left( \frac{z}{2-z}\right)^{2}  }
\end{equation}
For $z \in (-\infty,1]$, $t_*^-$ lies in the integration range $(0,1)$ while $t_*^+$ lies outside. Picking up the saddle $t_*^-$ the block at leading order in large $D$ is,
\begin{equation}\label{LDblockSP}
G_{\delta D,\ell}(z) =  \frac{1}{\sqrt{1-\left( \frac{z}{2-z}\right)^{2}}} A_{1-\ell}(1) \ \mathsf{v}_{\delta}(z) \ e^{D \hspace{.02cm} \mathsf{q}_{\delta}(z)}
\end{equation}
where,
%\begin{align}\label{qv}
%\mathsf{q}_{\delta}(z) &= \log \Big[ \sqrt{\frac{2(\Dd -1)^2 \hat{y}_{+}}{\left(2 \Dd -1\right) (B-2 \Dd +2 (\Dd -1) y_{+}+1)}} \left(\frac{2 (2 \Dd -1)^2 (B-2 \Dd +1) (B-2 \Dd +2 (\Dd -1) y_{+}+1)}{(1-B)\Dd (\Dd -1)^2   y_{+}}\right)^{\Dd /2} \Big] \notag \\
%\mathsf{v}_{\delta}(z) &= \sqrt{\frac{2^{-1}(\Dd -1)^{-3}(B-1)^3 \Dd  (2 \Dd -1) (B-2 \Dd +2 (\Dd -1) y_{+}+1)^2}{  \left(16 \Dd  (\delta -1)^2 (B-\Dd ) y_{+}^2 +16 \Dd  (\Dd -1)  (B-\Dd ) (B-2 \Dd +1) y_{+} + (B-2 \Dd +1)^2 \left(4 B \Dd +(B-1)^2-4 \Dd ^2\right) \right)}}
%\end{align}
\begin{align}\label{qv}
\mathsf{q}_{\delta}(z) &= \log \Big[ \sqrt{\frac{2(\Dd -1)^2 \hat{y}_{+}}{\left(2 \Dd -1\right) (B-2 \Dd +2 (\Dd -1) y_{+}+1)}} \left(\frac{2 (2 \Dd -1)^2 (B-2 \Dd +1) (B-2 \Dd +2 (\Dd -1) y_{+}+1)}{(1-B)\Dd (\Dd -1)^2   y_{+}}\right)^{\Dd /2} \Big] \notag \\
\mathsf{v}_{\delta}(z) &= \sqrt{\frac{(16 \delta)^{-1}(\Dd -1)^{-3}(B-1)^3 \Dd  (2 \Dd -1) (B-2 \Dd +2 (\Dd -1) y_{+}+1)^2}{   2  (\delta -1)^2 (B-4\Dd ) y_{+}^2 + 2(\delta-1) \left( B \left(1-3 \delta \right)+\left(8 \delta ^2-6 \delta +1\right) \right)y_{+} + (1 - 2 \delta)^2 (1 + B - 2 \delta)}}
\end{align}
with, $ B=\sqrt{1+4 (1-y_{+}) \Dd (\Dd -1)} $ and $ y_{+} = \left( \frac{z}{2-z}\right)^{2} $.\\
Given the large dimensional block \eqref{LDblockSP} in leading order in large $D$ limit, we will now compute the saddle point approximation of $\frac{\hat{G}_{\Delta,\ell}(z)}{z^{2 \Delta_{\phi}}}$ with respect to  $z$ in large $D$ limit. The saddle point equation and the solution for this is
\begin{equation}\label{saddleD}
\frac{d}{dz} \left( \mathsf{q}_{\delta}(z) -2 \delta_{\phi} \log (z) \right) =0 \qquad \Rightarrow \qquad z_* =    \frac{(2 \delta_{\phi} - \delta )( 2 \delta_{\phi} +\delta -1 )}{\delta_{\phi}(4 \delta_{\phi}-1)}.
\end{equation}
It is easy to verify that,
\begin{equation}
\frac{d^2}{dz^2} \left( \mathsf{q}_{\delta}(z) -2 \delta_{\phi} \log (z) \right) \big|_{z=z_*} \lessgtr 0 \qquad \textrm{for } \delta \gtrless 2 \delta_{\phi}
\end{equation}
This means that the path of steepest descent from the saddle point $z_*$ is along the real axis for $\delta> 2\delta_\phi$ while it is perpendicular to the real axis for $\delta<2\delta_\phi$. 

\bibliography{functional}

\end{document}